\documentclass[fleqn,usenatbib,twocolumn]{mnras}

\usepackage{newtxtext,newtxmath}

\usepackage[T1]{fontenc}
\usepackage{ae,aecompl}

\usepackage{etoolbox}
\makeatletter
\patchcmd\@combinedblfloats{\box\@outputbox}{\unvbox\@outputbox}{}{%
}%
\makeatother

\usepackage{graphicx}	
\usepackage{amsmath}	
\usepackage[usenames, dvipsnames]{color} 
\usepackage{float}
\usepackage{verbatim}
\usepackage{orcidlink}
\usepackage{xcolor}
\usepackage{supertabular}

\newcommand{\citesavard}{Savard et al. (submitted)}
\newcommand{\citepsavard}{(Savard et al., submitted)}
\newcommand{\citealtsavard}{Savard et al., submitted}

\title
[Blast waves and shocks in XRBs and GRBs]
{
Blast waves and reverse shocks: from ultra-relativistic GRBs to moderately relativistic X-ray binaries 
}

\author[J.~H. Matthews et al.]
{James H. Matthews$^{\orcidlink{0000-0002-3493-7737}}$,$^{1}$\thanks{james.matthews@physics.ox.ac.uk} Alex J. Cooper$^{\orcidlink{0000-0002-4033-3139},1}$, Lauren Rhodes$^{\orcidlink{0000-0003-2705-4941},2,3,1}$, Katherine Savard$^{\orcidlink{0009-0001-8598-0639},1}$, Rob Fender$^{1}$ \newauthor Francesco Carotenuto$^{\orcidlink{0000-0002-0426-3276},4}$, Fraser J. Cowie$^{\orcidlink{0009-0009-0079-2419},1}$, Emma L. Elley$^{\orcidlink{0009-0002-5349-908X},1}$, Joe Bright$^{\orcidlink{0000-0002-7735-5796},1}$, Andrew K. Hughes$^{\orcidlink{0000-0003-0764-0687},1}$ \newauthor 
and Sara E. Motta$^{\orcidlink{0000-0002-6154-5843},5,1}$
\\
$^{1}$Astrophysics Subdepartment, Department of Physics, University of Oxford, Keble Road, Oxford, OX13RH, UK\\
$^{2}$Trottier Space Institute at McGill, 3550 Rue University, Montreal, Quebec H3A 2A7, Canada\\
$^{3}$Department of Physics, McGill University, 3600 Rue University, Montreal, Quebec H3A 2T8, Canada\\
$^{4}$INAF-Osservatorio Astronomico di Roma, Via Frascati 33, I-00076, Monte Porzio Catone (RM), Italy\\
$^{5}$INAF-Osservatorio Astronomico di Brera, via E.\,Bianchi 46, 23807 Merate (LC), Italy
}

\date{Accepted 2025 April 9. Received 2025 April 7; in original form 2025 March 7}

\pubyear{2025}

\begin{document}
\label{firstpage}
\pagerange{\pageref{firstpage}--\pageref{lastpage}}
\maketitle

\begin{abstract}
Blast wave models are commonly used to model relativistic outflows from ultra-relativistic gamma-ray bursts (GRBs), but are also applied to lower Lorentz factor ejections from X-ray binaries (XRBs). Here we revisit the physics of blast waves and reverse shocks in these systems and explore the similarities and differences between the ultra-relativistic ($\Gamma \gg 1$) and moderately relativistic ($\Gamma \sim {\rm a~few}$) regimes. We first demonstrate that the evolution of the blast wave radius as a function of the observer frame time is recovered in the on-axis ultra-relativistic limit from a general energy and radius blast wave evolution, emphasizing that XRB ejections are off-axis, moderately relativistic cousins of GRB afterglows. We show that, for fixed blast wave or ejecta energy, reverse shocks cross the ejecta much later (earlier) on in the evolution for less (more) relativistic systems, and find that reverse shocks are much longer-lived in XRBs and off-axis GRBs compared to on-axis GRBs. Reverse shock crossing should thus typically finish after $\sim$10-100 of days (in the observer frame) in XRB ejections. This characteristic, together with their moderate Lorentz factors and resolvable core separations, makes XRB ejections unique laboratories for shock and particle acceleration physics. We discuss the impact of geometry and lateral spreading on our results, explore how to distinguish between different shock components, and comment on the implications for GRB and XRB environments. Additionally, we argue that identification of reverse shock signatures in XRBs could provide an independent constraint on the ejecta Lorentz factor. 
\end{abstract}


\section{Introduction}
Relativistic jets, and their associated blast waves and shocks, are ubiquitously produced in systems powered by accretion onto compact objects, from the ultra-relativistic flows produced in gamma-ray bursts (GRBs) to more mildly relativistic jet `ejections' in active galactic nuclei (AGN) and X-ray binaries (XRBs). The outflows are associated with transient explosive phenomena or persistent yet variable accretion on to a compact object. Monitoring of the radiation from the resulting blast waves provides important information about the energetics and nature of the underlying central engine. In addition, the blast waves transfer energy to the surrounding medium, accelerate high-energy particles, and act as a testbed for shock and plasma physics, often in a relativistic regime. 

A blast wave consists of a propagating shock front and associated region of high pressure produced by an impulsive injection of a large amount of energy. Much of our understanding of the kinematics of blast waves stems from analytic calculations of their behaviour in various limiting regimes. Supernova remnants are perhaps the most famous examples of astrophysical blast waves, and their adiabatic phase follows the well-known Sedov-Taylor-von Neumann self-similar solution \citep{taylor1950,sedov_examples_1958}. The ultra-relativistic analogue was developed by \cite{blandford_fluid_1976}, a self-similar solution which is thought to well-describe the relativistic phase of afterglows from gamma-ray bursts (GRBs). Extensions or variants of the Blandford-Mckee solution are frequently used to model the spectra and lightcurves of GRBs, typically based on analytic models for their evolution \citep{sari_spectra_1998,gao2013}. The evolution of the light curves and spectra changes depending on the structure of the jet \citep[whether it is Gaussian, top-hat, etc.;][]{ryan_afterglowpy2020} and changes of this structure over time \citep[whether it spreads laterally or not;][]{Panaitescu_1998, 1999ApJ...519L..17S}. However, it is often challenging for theoretical models to adequately fit the data (or for parameters to be well constrained) from the observed light curves and spectra alone \citep[e.g.][]{aksulu_exploring_2022}. In addition, the Blandford-Mckee and (early) GRB afterglow regimes are ultra-relativistic. Understanding the similarities and differences between the ultra-relativistic and mildly/trans-relativistic regimes of transients, jets, and blast waves is therefore important.

Direct information about the kinematics of blast waves in synchrotron-emitting transients, through the detection of proper motion, is invaluable to test the predictions of these theoretical models. In XRBs, blast wave-like phenomena can be produced during transient, large-scale ($\gtrsim 0.1$~pc), jet ejection events. The most famous example of these is the \cite{mirabel1994superluminal} discovery of superluminal apparent motion in the large-scale jet ejecta of GRS 1915+105. In recent years, the number of XRBs with newly detected, time-resolved, large scale jet ejection events has risen from three detected between 1998 and 2018 \citep{Hannikainen2001,gallo2004,yang2011}, to eight from 2018-2024 \citep[][with further work submitted or in prep.]{russell2019,bright_extremely_2020,carotenuto2021,bahramian2023}, albeit with the numbers depending on exactly how `large-scale' is defined. This order of magnitude increase in detection rate is principally due to dedicated monitoring with MeerKAT \citep{jonas_meerkat_2016} through the ThunderKAT program \citep{fender_thunderkat_2016} and its successor, X-KAT. 
The ejection events are linked to state transitions and radio flaring \citep{fender_towards_2004,fender2009}, and while their launching mechanism is not known, it is clear that a large amount of kinetic energy is released. The (presumably) kinetically-dominated ejecta then propagate into the ISM, decelerating and transferring energy and momentum to the surrounding environment. In the process, shocks are produced, which are thought to be the origin of the in-situ particle acceleration needed to explain the observed long-lived radio and X-ray emission \citep{bright_extremely_2020,espinasse2020}. The trajectories of the ejecta are tracked with regular radio monitoring and can be modelled using the same blast wave frameworks we will discuss in this work \citep{Carotenuto_2024}. This kinematic modelling can be used, in tandem with core monitoring in X-rays and radio, to infer information about both the central engine -- such as the total ejected energy, the connection to disc variability \citep{homan2020} and potentially even the composition of the ejected plasma \citep{zdziarski_cause_2024} -- and the ambient medium surrounding the XRB \citep{corbel_large-scale_2002,heinz2002,rushton_resolved_2017,Carotenuto_2022,Carotenuto_2024,sikora2023}. 

It is much more difficult to detect proper motion from GRBs; their angular separation from the core grows much more slowly with time, due to the cosmological distances to sources. High-resolution Very Long Baseline Interferometry (VLBI) provides a plausible pathway to direct kinematic measurements of these objects. Two events have associated ``expansion speed'' measurements: GRB 030329A and 221009A, \citep{2004ApJ...609L...1T, 2024A&A...690A..74G}, while two further studies obtain upper limits using similar techniques (GRB 201015A, 
 \citealt{2022A&A...664A..36G}; GRB 190289A, \citealt{Salafia_2022}). Moreover, the proximity of GRB170817a \citep{ligo_mm_2017}, associated with the seminal gravitational wave event GW170817 \citep{ligo_gw_2017}, enabled a measurement of jet proper motion at both radio \citep{mooley_radio_18} and optical \citep{mooley_optical_2022} wavelengths. GRB proper motions from the aforementioned studies provide additional constraints for afterglow modelling \citep[e.g.][]{ryan2024}, inform the study of multiple shock components, and provide additional constraints on the circumburst medium.  

Both XRBs and GRBs represent rich laboratories for relativistic astrophysics, particularly concerning shocks and particle acceleration. GRBs are thought to have multiple shock components: a forward shock accelerates particles throughout the afterglow\footnote{GRBs are characterised by an initial burst of gamma-rays on timescales ranging from $<1$s to $\sim 1000$s \citep[the prompt emission; e.g.][]{prompt_zhang}, followed by a much longer-lived afterglow phase in which the remaining energy associated with the GRB is gradually dissipated \citep[e.g.][]{piran1999}.} phase \citep{meszaros1997,sari_spectra_1998,piran1999,wang_how_2015}, and internal shocks plausibly power the prompt emission \citep{rees_unsteady_1994}. In addition, at fairly early times ($\sim$days), it is sometimes possible to detect emission associated with a reverse shock component \citep{laskar2013,laskar2016,laskar2019,lamb2019,2020MNRAS.496.3326R,bright_rhodes_2023,rhodes2024}. In XRBs, the role of shocks is perhaps less well established, but a qualitatively similar picture can be painted in which internal shocks are important in the `compact jet' phase \citep{kaiser_internal_2000,malzac_spectral_2014,malzac_jet_2018} and radio flare \citep{fender2009}, with the radio emission from the large scale jet ejecta being powered by {\sl in situ}  particle acceleration at a forward or external shock \citep{corbel_large-scale_2002,rushton_resolved_2017,bright_extremely_2020}. Reverse shocks have also been discussed in the large-scale decelerating jets considered here \citep[][\citealtsavard]{wang_external_2003,Hao}, but their importance remains somewhat unclear. 

In both source classes, the shock structures are traced through the non-thermal particles that they accelerate. Partly as a result of this, XRB and GRB jetted blast waves have often been discussed as potential sources of cosmic rays (CRs). In particular, XRB jets may represent a contribution to Galactic CRs up to and perhaps beyond the `knee' at $\sim$PeV energies \citep{heinz_cr_2002,fender_energization_2005,cooper_high-energy_2020,Kantzas2023}, an idea that has been bolstered by the recent detection of very-high energy gamma-rays from a host of Galactic XRBs or microquasars \citep{Abeysekara2018,lhaaso2024,alfaro2024}. TeV gamma-ray detections have also been reported from GRBs, notably from GRB 221009A \citep{cao_very_2023}. However, given their prodigious powers, GRBs may be able to accelerate particles to much higher energies, and they are potential sources \citep{waxman_cosmological_1995,vietri_acceleration_1995,baerwald_are_2015,globus_uhecr_2015} for ultra-high energy cosmic rays (UHECRs), protons and nuclei with energies up to $10^{20}$~eV \citep[see][for recent UHECR reviews]{alves_batista_open_2019,matthews_review_2023,biteau2024}. Important theoretical questions persist regarding particle acceleration in relativistic shocks \citep{lemoine_efficiency_2006,sironi_particle_2011,sironi_maximum_2013,reville_maximum_2014,marcowith_microphysics_2016,ellison_particle_2016,matthews_particle_2020,huang_prospects_2023}, in particular how the total amount of energy given to nonthermal particles, and their maximum energies, depend on shock parameters like Lorentz factor and magnetisation. Both GRBs and XRBs represent important observational testbeds for theoretical and numerical studies of these topics; there are also additional synergies with other jetted systems such as tidal disruption events (TDEs), pulsar wind nebulae, and AGN. 

Despite the clear similarities between GRB and XRB phenomenology -- and their complementary nature as probes of important and interesting physics -- the extent to which they can be unified is not always clear, and their literature trails are somewhat divergent. One obfuscating factor is that they lie in different regimes in the relevant energy versus viewing angle versus Lorentz factor parameter space. However, there are also very real differences in their evolution. In this paper, we aim to discuss and address some of these issues, and try to answer the following questions: Are GRB afterglows and XRB jet ejections `cousins', that is, similar phenomena in different regimes? What is the role of reverse shocks? How do the underlying assumptions in blast wave models affect the inferred parameters? Does it make sense to apply GRB blast wave models to XRB ejecta? We start by providing the observational context in section~\ref{sec:context}, before describing the blast wave models and kinematics in section~\ref{sec:blast-waves}. In section~\ref{sec:critical}, we identify the critical radii in the evolution of the blast wave and explore the dependence on Lorentz factor. We discuss the associated observed timescales and the overall evolution of the blast wave system in section~\ref{sec:timescales}. In section~\ref{sec:geometry}, we explore how different geometries and lateral spreading may affect the propagation physics, before discussing our results in section~\ref{sec:discuss}, focusing on ambient densities, constraining the initial Lorentz factor and using XRBs as shock laboratories. Finally, in section~\ref{sec:conclusions}, we summarise our results and conclude.

\section{Observational Context}
\label{sec:context}
Before discussing blast wave models, we set the scene by considering the observational constraints on the key quantities of jet ejections and relativistic blast waves. Specifically, we first seek constraints on the total energy and opening angles of blast waves driven by GRBs and XRB transient jet ejections. In addition, we collate a set of (model-dependent) estimates of Lorentz factors of the same systems. We discuss instantaneous estimates of Lorentz factor $\Gamma$, as well as inferred initial Lorentz factors, $\Gamma_0$, which is the value of $\Gamma$ at the start of the propagation of a discrete ejection or blast wave shell (neglecting any initial acceleration period or more complex launch dynamics). We also define the usual $\beta$, the bulk velocity in units of $c$, and $\beta_{\rm app}$, the apparent velocity.

\begin{figure*}
    \centering
    \includegraphics[width=\linewidth]{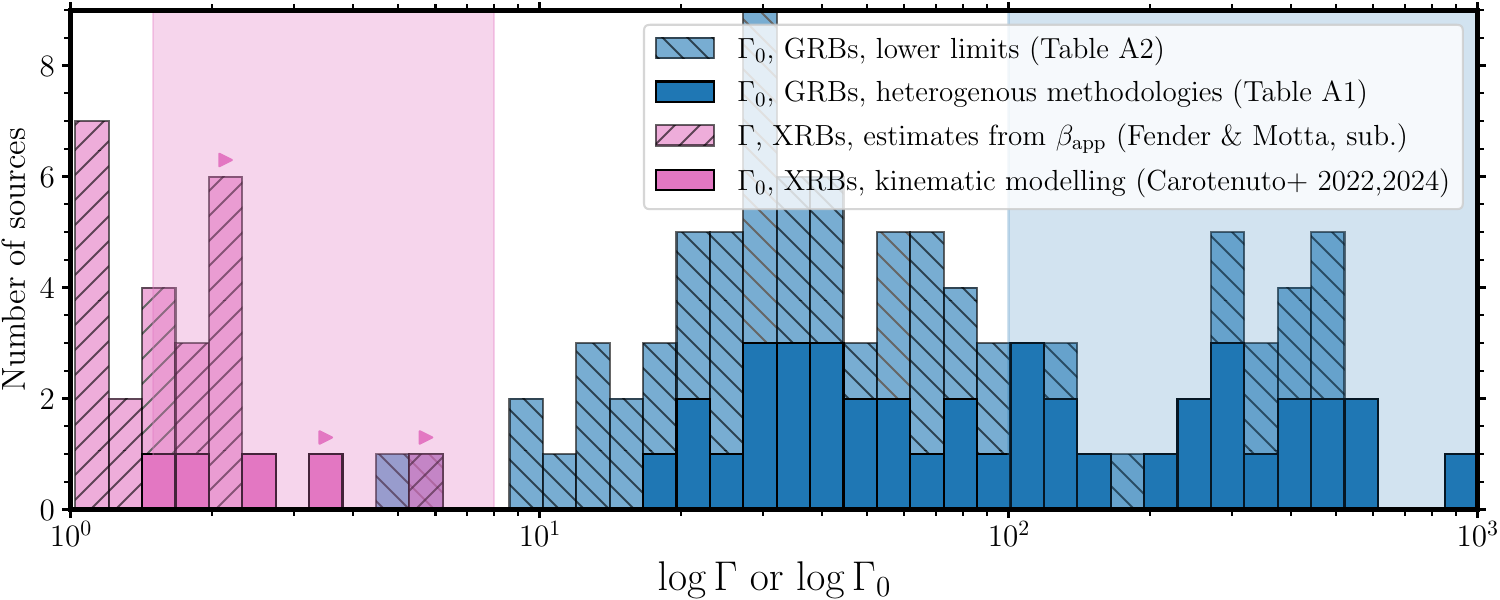}
    \caption{
    Stacked histograms of Lorentz factor estimates in XRBs (pink) and GRBs (blue), with different methods and data sources labelled. Lower
    limits for XRBs are marked with a pink triangle. With various caveats (see text), GRBs blast waves are ultra-relativistic, and at least some XRB jet ejecta are mildly or moderately relativistic. The banded pink and blue regions show the approximate regimes used to aid our discussion of XRB ($1.5\lesssim \Gamma_0 \lesssim 8$) and GRB ($\Gamma_0 \gtrsim 100$) blast waves, respectively. 
    }
    \label{fig:lorentz}
\end{figure*}

\subsection{X-ray binaries}
In XRBs, there are multiple ways to constrain the properties of both transient jets and the persistent, compact jets seen in the hard spectral state. The method most directly relevant to our work is the application of the external shock model to large-scale decelerating transient jets \citep[see section~\ref{sec:blast-wave-models};][]{wang_external_2003,Steiner_xte,Carotenuto_2022,Carotenuto_2024,zdziarski2023}, which can be used to constrain both the Lorentz factor, and the effective initial energy. The latter is given by $\bar{E}_0 = E_0 / (n_{\rm ism} \phi_\circ^2)$, required due to an inherent degeneracy between the three parameters: $E_0$, the initial energy, $n_{\rm ism}$, the ISM density in ${\rm cm}^{-3}$ and $\phi_\circ$, the half-opening angle in degrees. 
\cite{Carotenuto_2024} infer effective energies ranging from $\bar{E}_0\approx10^{45}~{\rm erg}$ for XTE J$1752-223$ up to $\bar{E}_0\approx6\times 10^{48}~{\rm erg}$ for MAXI J$1535-571$. Such high effective energies strongly suggest a significantly underdense ISM compared to the canonical value of $1$~particle~cm$^{-3}$, as has been discussed by various authors \citep[][\citealtsavard]{heinz2002,Carotenuto_2022,Carotenuto_2024,zdziarski2023,zdziarski_cause_2024}. Even with an underdense ISM the energies released are significant; for example, in MAXI J1820$+$070 at 90 days post launch, \cite{bright_extremely_2020} find the internal energy alone must be at least $10^{41}~{\rm erg}$, and the total energy is likely to be significantly higher than this conservative estimate. 
 
The Lorentz factors of transient XRB jet ejections are challenging to constrain or measure, even when the distance to the source is known. The principal difficulty is that, for a significantly off-axis source, an observed superluminal apparent speed is entirely consistent with almost any Lorentz factor greater than 1; to be concrete, for a typical XRB viewing angle of $60^\circ$ then a blob or jet ejection with $\Gamma = 2$ propagates with almost an identical proper motion to one with $\Gamma = 1000$. This degeneracy means that the Lorentz factor can only really be constrained from proper motions alone for the relatively few sources that are viewed at low inclination, or those that have slower propagation speeds. Two such examples are 4U 1543--47, a low inclination source with an estimate of $\Gamma_0 \approx 8$ (Zhang et al., submitted), and MAXI J1848-015, which has an ejecta velocity of $\approx0.8 c$ \citep[$\Gamma\approx 1.67$;][]{bahramian2023}.

Alternatively, the Lorentz factor can sometimes be estimated by fitting the trajectories of both the approaching and receding components. As part of their Bayesian fitting framework \cite{Carotenuto_2022,Carotenuto_2024} obtain constraints of $\Gamma_0 = 1.85^{+0.15}_{-0.12}$ for MAXI J1348-630, $\Gamma_0 = 2.6^{+0.5}_{-0.4}$ for MAXI J1820$+$070, $\Gamma_0 = 1.6 \pm {0.2}$ for MAXI J1535-571, and $\Gamma_0 > 3.4$ for XTE J1752-223. Thus, overall, while XRB jet ejections are clearly not ultra-relativistic, there is still considerable uncertainty over their distribution of initial Lorentz factors $\Gamma_0$; all we can be sure of is that at least a few are moderately relativistic ($\Gamma_0 \sim$ a few), and some are likely $\sim 10$. We note that there is also indirect evidence that some (neutron star) XRBs produce $\Gamma > 10$ outflows \citep{fender2004_urf,motta2019}, but these are not our focus. 

\subsection{Gamma-ray bursts}
GRB Lorentz factors early in the burst must, in general, be high ($\Gamma \gtrsim 100$), else the optical depths to gamma-ray photons would be too high for them to be observed \citep[{\rm the compactness problem}, see e.g.][]{piran1999}. \cite{Lithwick2001} derive a secondary limit, which is sometimes more stringent, based on scattering off $e^- e^+$ pairs. More generally, the relativistic fireball model \citep{rees1992,Meszaros1993,piran1999} is widely accepted as explaining the overall phenomenology of GRBs (albeit with ongoing debates about jet structure, shocks, prompt emission mechanisms, etc.). 

However, while it is clear that GRBs must be highly relativistic, obtaining actual quantitative estimates of the initial Lorentz factor $\Gamma_0$ of the afterglow blast wave  -- and knowing whether Lorentz factors obtained from, e.g., opacity limits correspond directly to $\Gamma_0$ -- is a thorny subject.  With very few proper motion constraints (see section~\ref{sec:discuss_slowing}), estimates of the initial Lorentz factor come from model-dependent or indirect means. For example, a typical method is to study the early afterglow and search for reverse and forward shock peak times \citep[e.g.][]{zhang2003} or, related, so-called ``deceleration signatures'' or ``onset bumps'' in which the peak time of the optical or gamma-ray light curve depends sensitively on $\Gamma_0$ \citep{Ghirlanda2018}. \cite{Liang2010} use this onset bump method to measure initial Lorentz factors of $\sim$hundreds and a tight relationship between $\Gamma_0$ and $E_{\gamma,{\rm iso}}$ in a sample of $19$ GRBs, where $E_{\gamma,{\rm iso}}$ is the isotropic equivalent gamma-ray energy released. Similarly, \cite{Ghirlanda2018} provide measurements of $\Gamma_0$ from 67 GRBs with peaks in their optical or GeV light curves. 
Finally, \cite{peer2007} propose a method based on measuring the temperature and flux of the thermal component seen at early times in gamma-rays and X-rays. 

$E_{\gamma,{\rm iso}}$ can be measured directly from the burst fluence if the GRB redshift is known, with typical values in the range $10^{52-54}~{\rm erg}$ \citep{frail2001,amati2006,Atteia2017}. There is evidence for a cutoff in the distribution at $\log_{10} E_{\gamma,{\rm iso}} \sim 54.5$ \citep{Atteia2017, 2022ApJ...940L...4D}. GRB 221009A, with $E_{\gamma,{\rm iso}} \sim 55$ \citep{lesage2023,An2023,Frederiks2023}, is a notable extremely energetic burst beyond (but broadly statistically consistent with) this cutoff \citep{Atteia2025}. Considering that the opening angle of GRB jets are likely a few degrees \citep{frail2001, 2015ApJ...815..102F}, this means the true radiated energy is 100-1000 lower than $E_{\gamma,{\rm iso}}$, of order $10^{51}~{\rm erg}^{-1}$ and comparable with the energy released by a core-collapse supernovae \citep{frail2001,Woosley2006,Golstein2016}.

The prompt emission does not represent the entire energy budget of the GRB; a significant fraction, and probably the majority, of the total energy is made up of kinetic energy that is dissipated gradually through the resulting blast wave and synchrotron afterglow -- this kinetic energy is the relevant energy reservoir for our work. An important parameter is therefore the radiative efficiency, $\epsilon_\gamma \equiv E_{\gamma,{\rm iso}} / (E_{\gamma,{\rm iso}}+E_{k,{\rm iso}})$, where we have introduced $E_{k,{\rm iso}}$, the isotropic equivalent kinetic energy. Measurements of $E_{k,{\rm iso}}$ are challenging, but in the absence of fully multi-wavelength afterglow modelling, the X-ray afterglow is thought to act as an effective probe of the kinetic energy budget \citep{freedman2001,2004ApJ...613..477L,fan2006,zhang2007}. Typically, radiative efficiencies of $\sim 0.1$ are inferred from these methods, a value we use to inform our choice of blast wave energy for the GRB case, although we note some studies find much lower values \citep[e.g.][]{Salafia_2022}.

\subsection{Collating estimates and representative values}
\label{sec:represent}
Given all the above uncertainties, it is difficult to choose representative values for the initial kinetic energy, $E_0$, initial Lorentz factor, $\Gamma_0$, ISM number density, $n_{\rm ism}$, and jet/outflow opening angle, $\phi$. Where relevant, we assume the blast waves are travelling into an relatively low density ISM (see section~\ref{sec:discuss_slowing}) with number density $n_0=6.95 \times 10^{-3}~{\rm cm}^{-3}$ such that the mass density is $\rho_0 = n_0~m_p$.
We adopt representative values of $\Gamma_0 = 2.5$, $E_0 = 10^{44}~{\rm erg}$, and $\phi = 1^\circ$ for XRB jet ejections, and $\Gamma_0 = 100$, $E_0 = 10^{51}~{\rm erg}$, and $\phi = 1^\circ$ for GRBs, noting that for $\phi = 1^\circ$ the isotropic equivalent kinetic energy is $E_{k, {\rm iso}} = (180 / \pi)^2 E_0 \approx 3282 E_0$. Our adopted GRB half-opening angle is a little narrow; they are typically a few degrees and sometimes larger \citep{frail2001,racusin2009,Ghirlanda2013,2015ApJ...815..102F,ryan2015,Golstein2016}. However, adopting the same value for XRBs and GRBs makes the scaling of results more straightforward for the reader. Furthermore, within the external shock model we will discuss (section~\ref{sec:blast-wave-models}), $n_0$ and $\phi^2$ are both degenerate with the choice of $E_0$ and in many cases will be normalised away, and it is also straightforward to scale the results to an alternative choice of $n_0$ or $\phi$.

In many cases, we will plot quantities normalised in such a way that neither the energy, opening angle nor the ISM density appears explicitly or matters. We stress that these values are illustrative and should be thought of as delineating approximate physical regimes, since the properties of GRB and XRB jets at the population level are somewhat diverse and poorly constrained. 

To illustrate this latter point, we collate some estimates of $\Gamma_0$ and $\Gamma$ -- {\em all} of which are model-dependent -- from the literature for both GRBs and XRBs. These estimates are plotted as histograms in Fig.~\ref{fig:lorentz}. Our XRB measurements are obtained from the two methods described above: we take estimates of initial Lorentz factors, $\Gamma_0$, for four XRBs from \cite{Carotenuto_2022} and \cite{Carotenuto_2024}, and instantaneous Lorentz estimates from the apparent speeds collated by Fender \& Motta (submitted). The GRB measurements come from a more diverse set of publications, which are given in Appendix~\ref{sec:grb_table} (Tables~\ref{grb_table1} and \ref{grb_table2}). We do not include errorbars or attempt to ensure a consistent statistical methodology: we therefore stress again that these Lorentz factors estimates are heterogeneous, approximate and subject to a wide range of selection effects. Hereafter, we focus on moderately relativistic XRB ejecta with $\Gamma_0 \sim 1.5-8$, and highly relativistic GRB ejecta with $\Gamma_0 \gtrsim 100$, but that is not to say that all sources will produce jets or blast waves with $\Gamma_0$ lying in this range \citep[see, e.g.][for a particularly relevant GRB study]{dereli2022}. 
These Lorentz factor ranges are illustrated in Fig.~\ref{fig:lorentz}, and throughout, with shaded regions. 

\begin{figure}
\centering
\includegraphics[width=\linewidth]{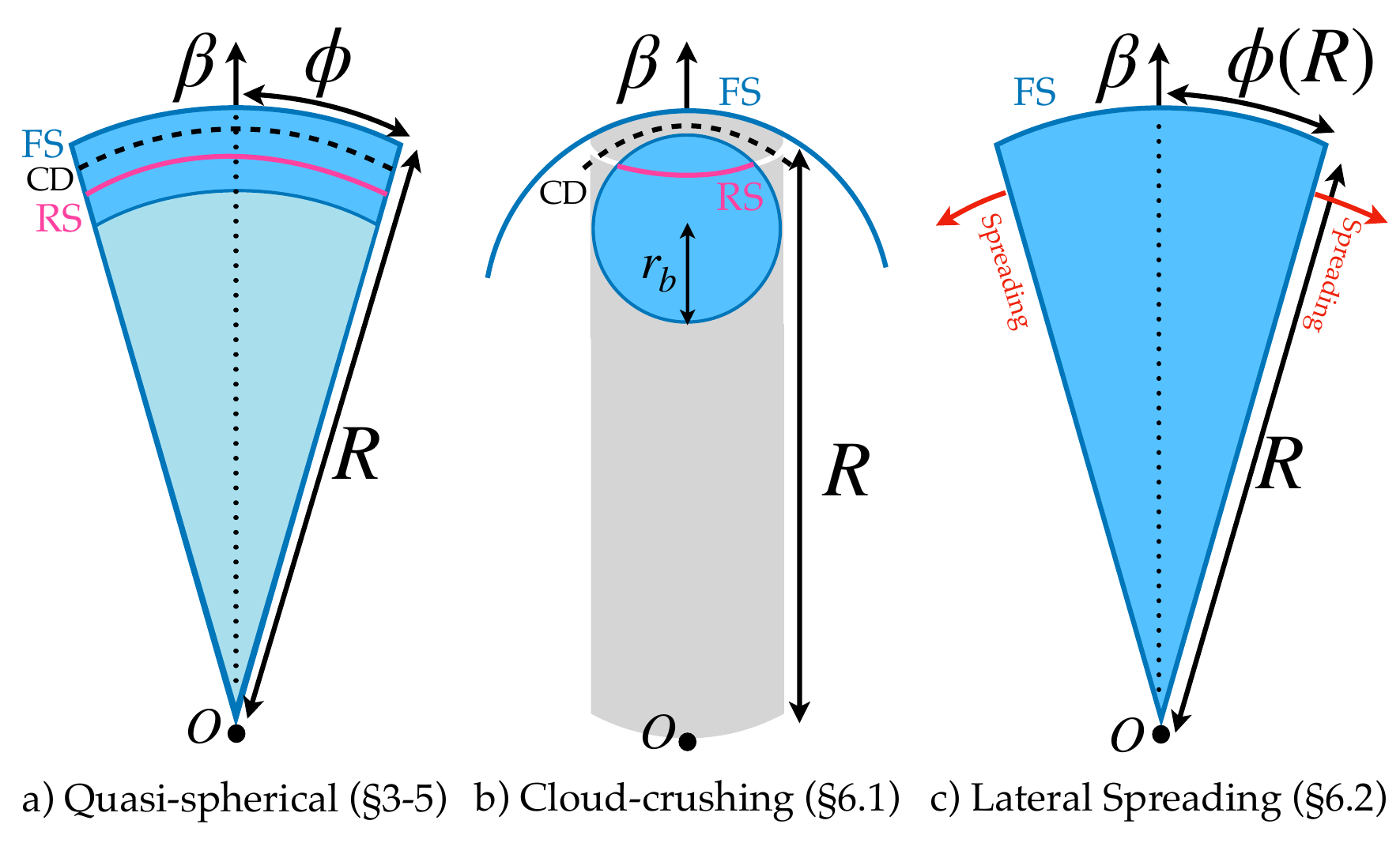}
\caption{Schematic diagram showing the geometries considered in this paper. $R$ is the laboratory frame distance from the origin ($O$) to the forward shock (FS). The reverse shock (RS; pink sold line) and contact discontinuity (CD; dashed black line) are labelled.    
Initially, we focus purely on the quasi-spherical or conical geometry shown in panel a). We discuss the alternative geometries shown in panels b) and c) in section~\ref{sec:geometry} as well as the overall impact of geometry on the propagation of blast waves and associated inference of physical parameters. The reverse shock or contact discontinuity are not shown in panel c) because spreading occurs only after the two-shock phase (see section~\ref{sec:discuss_lateral}).   
}
\label{fig:geometry}
\end{figure}

\section{Blast wave models and kinematics}
\label{sec:blast-waves}
We will proceed by briefly reviewing some commonly used blast wave models in the literature and discuss the various assumptions and limits in which each can be applied. Throughout this section we will assume a quasi-spherical blast wave with a solid angle $\Omega$ (panel a) of Fig.~\ref{fig:geometry}); this assumption is interrogated in Section~\ref{sec:geometry}. For an outflow with a spherical cap or conical geometry, the solid angle is related to the half-opening angle $\phi$ via $\Omega = 2 \pi (1- \cos \phi)$, which is $\approx \pi \phi^2$ for small $\phi$. Although the external shock model described in section~\ref{sec:blast-wave-models} has a defined half-opening angle, it is quasi-1D in nature and so does not account for angular or lateral structure as one moves away from the axis of propagation. As a result, here we take on-axis to mean {\em strictly} on-axis, i.e. with viewing angle $\theta=0$, but this does not mean our on-axis results are limited to that case; rather, they are broadly applicable to situations when $\theta < \phi$. This is important to note given that GRBs are thought to be typically observed somewhat off-axis but still satisfying this condition \citep{ryan2015}.

For time evolution, we consider $t$ as the time as measured in the observer frame (although neglecting the effects of cosmological redshift for GRBs), which includes time dilation and geometric light travel time effects and thus depends on viewing angle $\theta$ \citep[see e.g.][ for further discussion]{zhang2004,zhang2009}. $R$ is the laboratory frame distance from the origin of the blast wave, which can be converted to measured projected angular distance, $\alpha$, again ignoring cosmological effects, through the formula $\alpha = R \sin \theta / D$, where $D$ is the distance to the source. 

\subsection{Ultra-relativistic blast waves and the Blandford-Mckee model}
\label{sec:blandford-mckee}
\defcitealias{blandford_fluid_1976}{BM76}
In the Blandford-Mckee model \citep[][hereafter \citetalias{blandford_fluid_1976}]{blandford_fluid_1976}, the explosion after the reverse shock phase is modelled as a self-similar, expanding spherical blast wave in which the total energy is conserved. In this solution, the density profile is usually assumed to follow $\rho(R) = \rho_0 R^{-k}$, with canonical profiles typically considered in GRB afterglow modelling corresponding to wind-like ($k=2$) and uniform density interstellar medium (ISM; $k=0$). We will adopt $k=0$ for simplicity such that $\rho(R)=\rho_0$, but our results can easily be generalised to $k\neq0$ (see, e.g., section~\ref{sec:critical} and \citealt{piran1999}). The energy of a relativistic blast wave sweeping up matter from the ISM is given by 
\begin{equation}
E =  \frac{\Omega}{3} \Gamma^2 \rho_0 c^2 R^3,
\label{eq:grb_energy}
\end{equation}
If we now enforce energy conservation, such that $E=E_0$ with $E_0$ the initial blast wave energy, this implies the relationship $R \propto \Gamma^{-2/3}$. It is common to introduce the `isotropic equivalent energy', that is, the total energy of the blast wave if $\Omega = 4\pi$, given by $E_{k,{\rm iso}} = 4\pi E_0 / \Omega$. It is convenient to introduce a generalized Sedov-Taylor-von Neumann scale,
\begin{equation}
l_{S}=\left( \frac{3E_0}{\rho_0 c^2} \right)^{1/3},
\label{eq:sedov_noomega}
\end{equation}
and the quasi-spherical version of this, including a correction for opening angle:
\begin{equation}
l =\left( \frac{3E_0}{\Omega \rho_0 c^2} \right)^{1/3} = \left( \frac{3E_{k,{\rm iso}}}{4\pi \rho_0 c^2} \right)^{1/3} .
\label{eq:sedov}
\end{equation}
The physical meaning of the Sedov-Taylor-von Neumann length is discussed further in section~\ref{sec:critical}. Using these quantities and applying energy conservation, one can show that the evolution of $R$ and $\Gamma$ over time is then \citep[e.g.][]{piran1999}
\begin{equation}
 R(t) =  (2 l^{3})^{1/4} t^{1/4},
 \end{equation}
\begin{equation}
 \Gamma(t) = (l^{3} / 8)^{1/8}
 t^{-3/8} \, .
\end{equation}
Now following geometric arguments, in an ultra-relativistic blast wave, photons emitted while the shell moves a distance $dR$ arrive, for an on-axis observer, on a timescale $dt\approx dR/(2 \Gamma^2 c)$ \citep{sari1997,waxman1997}, or alternatively
\begin{equation}
\frac{dR}{dt} \approx 2 \Gamma^2 c \, .
\label{eq:drdt_grb}
\end{equation}

\subsection{The external shock model}
\label{sec:blast-wave-models}
We now introduce the generalised blast wave or `external shock' model, which is no longer limited to the ultra-relativistic or on-axis cases. The external shock model is described by \cite{Huang_1999} and first applied to XRBs by \cite{wang_external_2003}. Subsequently, a number of groups have applied the model to fit the proper motions of discrete jet ejecta in XRBs \citep{Hao,Steiner_xte,steiner_2012b,Carotenuto_2022, Carotenuto_2024,zdziarski2023}. A related model is described by \cite{peer2012} who explores the behaviour in both the GRB and XRB regimes. We consider a quasi-spherical, thin shell of ejecta with initial mass $M_0$ and initial Lorentz factor $\Gamma_0$. The initial kinetic energy of the shell is 
\begin{equation}
E_0 = (\Gamma_0 - 1) M_0 c^2 \, .
\label{eq:kinetic_energy}
\end{equation}
The general blast wave model is based on energy conservation such that total energy change $dE/dt = 0$ and the energy lost from the ejecta is transferred to the swept up mass, that is
\begin{equation}
E_0 = (\Gamma - 1) M_0 c^2 + \sigma (\Gamma_{\rm sh}^2 - 1) m_{\rm sw} c^2 \, ,
\label{eq:energy_cons}
\end{equation}
where $\Gamma_{\rm sh}$ is the Lorentz factor of the (forward) shocked material, $\sigma$ is a numerical factor equal to $6/17 (\approx 0.35)$ for ultra-relativistic shocks and $\approx0.73$ for non-relativistic shocks. $\Gamma$ and $\Gamma_{\rm sh}$ are related as \citepalias{blandford_fluid_1976}
\begin{equation}
\Gamma_{\rm sh} = 
\frac{(\Gamma + 1) (\hat{\gamma} (\Gamma - 1) +1)^2}
{\hat{\gamma} (2-\hat{\gamma})(\Gamma - 1) + 2}
\label{eq:gammash}
\end{equation}
where $\hat{\gamma}$ is the adiabatic index and we adopt the formula of \cite{Steiner_xte} and \cite{Carotenuto_2022,Carotenuto_2024} for interpolating between the non-relativistic ($\hat{\gamma} = 5/3$) and ultra-relativistic ($\hat{\gamma} = 4/3$) regimes as $\hat{\gamma}=(4\Gamma+1)/(3\Gamma)$. Equation~\ref{eq:gammash} asymptotes to the ultra-relativistic and non-relativistic jump conditions of $\Gamma = \sqrt{2}\Gamma_{\rm sh}$ \citep[e.g][]{achterberg_particle_2001} and $\beta = 4\beta_{\rm sh}/3$ \citep[e.g][]{landau1959}.

We assume the outflow occupies a solid angle $\Omega$ such that the swept up mass obeys 
\begin{equation}
m_{\rm sw} = \frac{\Omega \rho_0 R^3}{3} \, .
\label{eq:mass_swept}
\end{equation}
This evolution is solved numerically by \cite{Carotenuto_2024} and others to obtain the solution for blast wave radius over time -- the change in (deprojected) radius for time measured in the observer frame follows the relation
\begin{equation}
\frac{dR}{dt} = \frac{\beta c}{1 \mp \beta \cos \theta} \, ,
\label{eq:drdt_xrb} 
\end{equation}
where the $\mp$ denotes approaching and receding components, respectively. 
This equation holds in general, that is, for varying $\beta$ and all $\theta$. The angular separation as a function of observer time, $\alpha(t)$, can then be calculated as 
\begin{equation}
\alpha(t) = D^{-1} \int_0^t \frac{\beta(t^\prime) c}{1 \mp \beta(t^\prime) \cos \theta} \sin\theta ~dt^\prime \, ,
\label{eq:alpha_xrb} 
\end{equation}
which now includes the $\sin\theta$ factor for the projected distance. One can then numerically obtain the solution and fit it to observations of $\alpha(t)$, for parameters $\Gamma_0$, $\Omega$ and $E_0$, with a degeneracy between $\Omega \sim \phi^2$, $E_0$, and $\rho$. To implement the above model, we use a modified version of the \texttt{JetKinematics}\footnote{\url{https://github.com/f-carotenuto/JetKinematics}} code presented by \cite{Carotenuto_2024}. 

\begin{figure}
\centering
\includegraphics[width=\columnwidth]{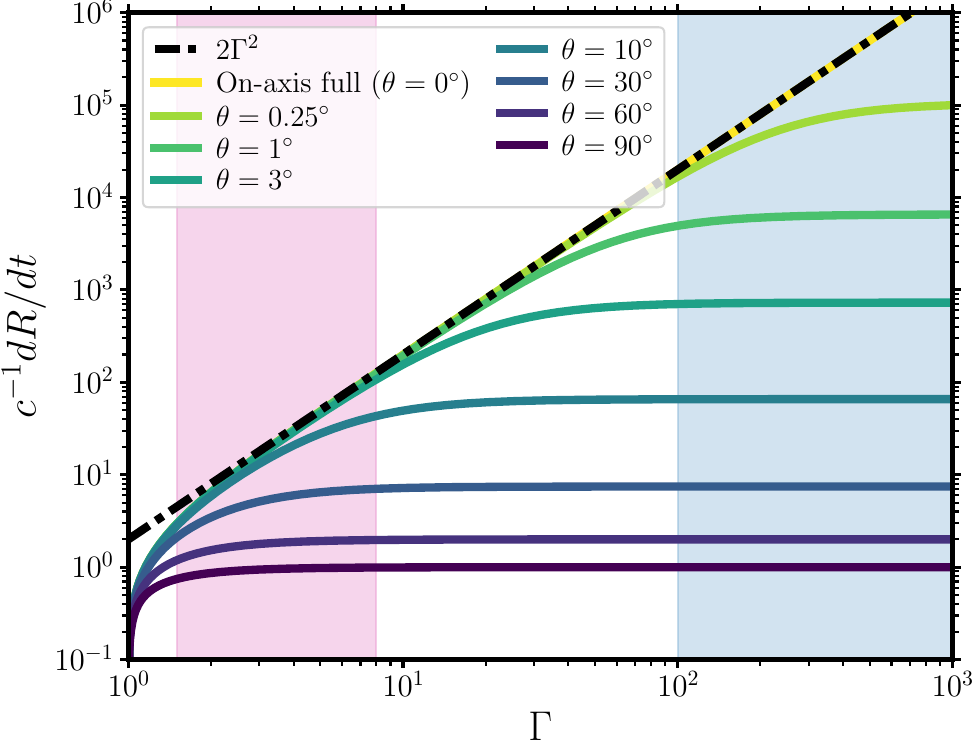}
\caption{The dependence of  the rate of change of blast wave radius with respect to observer time, $dR/dt$, on Lorentz factor $\Gamma$ and viewing angle $\theta$, for ejecta approaching the observer. The coloured lines show the full equation~\ref{eq:drdt_xrb}, compared with the on-axis limit from equation~\ref{eq:drdt_grb} shown with a dot-dashed line. The pink and blue shaded regions show approximate initial Lorentz factor ranges for XRBs and GRBs. At $\theta=0^\circ$, the solution agrees with the $2 \Gamma^2$ asymptote for high $\Gamma$. At small angles to the line of sight, the curves also agree with this curve for intermediate $\Gamma$ before flattening off as the $\cos \theta$ term becomes important. 
}
\label{fig:dR_dt}
\end{figure}

\subsection{On-axis relativistic blast waves}
\label{sec:on-axis}
Here we show that equation \ref{eq:drdt_grb} can be obtained by taking the ultra-relativistic, on-axis limit of equation~\ref{eq:drdt_xrb}. This derivation is very similar to the `textbook result' that the Doppler factor goes as $2 \Gamma^2$ in the on-axis, ultra-relativistic case \citep[see also section 3.2 of][]{zhang2004}; nevertheless, for completeness and clarity, we include it here. 

We begin by making the {\em on-axis} assumption, i.e. that $\cos \theta \approx 1$ and the ejecta is moving towards us, so that 
\begin{equation}
\frac{dR}{dt} = \frac{\beta c}{1 - \beta}
\label{eq:drdt}
\end{equation}
multiplying top and bottom by $1+\beta$ gives 
\begin{equation}
\frac{dR}{dt} = c \frac{\beta (1+\beta)}{(1 - \beta)(1+\beta)} = c \frac{\beta + \beta^2}{1-\beta^2}
\end{equation}
we now make the substitution $\beta = \sqrt{1-\Gamma^{-2}}$ which, after some manipulation and noting that  $1-\beta^2 = \Gamma^{-2}$, gives
\begin{equation}
\frac{dR}{dt} = c \left[\frac{\sqrt{1-\Gamma^{-2}}}{\Gamma^{-2}} + \Gamma^2 + 1\right] .
\end{equation}
Finally, we take the {\em ultra-relativistic limit} $\Gamma\to \infty$: as $\Gamma\to \infty$, $\Gamma^{-2}\to 0$ and the first term tends to $\Gamma^2$. As a result, we obtain $dR/dt \approx 2 \Gamma^2 c$, reproducing the dependence above. We therefore find that the equations for the evolution of $R$ over time are the ultra-relativistic, on-axis limit of equation~\ref{eq:drdt_xrb}: $dR/dt \propto \Gamma^2 c$ should hold for {\em any} $\Gamma(t)$ if $\Gamma \gg 1$ and $\theta \sim 0$, and $R(t) \propto t^{1/4}$ describes the specific evolution for  $\Gamma(t)\propto t^{-3/8}$. 

A graphical representation of the above derivation is shown in Fig.~\ref{fig:dR_dt}, which shows the recovery of $dR/dt \approx 2\Gamma^2 c$ behaviour in the GRB (on-axis, ultra-relativistic) limit. We show a variety of viewing angles in this figure, designed to illustrate a few interesting behaviours. The small $\theta$ curves show the $\Gamma^2$ dependence for intermediate $\Gamma$, but for high $\Gamma$ the $\cos \theta$ term in the denominator of equation~\ref{eq:drdt} becomes important and the curves flatten off. For larger viewing angles, the curve never intersects the $dR/dt=2\Gamma^2$, instead turning over and flattening off, with the $\theta=90^\circ$ curve purely following the relation $dR/dt= \beta = \sqrt{1-\Gamma^{-2}}$. 

\subsection{Terminology}
Before proceeding further, we wish to make a few points on terminology. First, we note the potential for confusion regarding what is meant by {\sl deceleration}. Deceleration could be used to refer to an observable decrease in the rate of change of angular separation over time. Other definitions could relate to the point at which the blast wave {\sl starts} deceleration, or when the blast wave sweeps up a rest mass equal to $\Gamma^{-1}$ of the initial mass, which is often how deceleration time, $t_{\rm dec}$, is used in the GRB literature. To avoid conflation of these (and other) possible meanings, we tend to avoid the use of the term and refer explicitly to  well-defined critical radii in the next section. 

In addition, a potential area of confusion when interpreting observations in the blast wave model framework is in the distinction between the distance travelled by the ejecta from the central engine ($R$) and the size of the emitting region. Often estimations of the size of the emitting region from techniques such as synchrotron self-absorption measurements or interstellar scintillation are used as a measurement of $R$. However, we stress that the size of the emission region and $R$ are two different quantities, and while related for some geometries (such as a conical one), this is not always the case. Equivalently, the expansion velocity of the emitting region and the velocity of the ejecta/blast wave are not equivalent or even necessarily related in a simple fashion. Therefore, we caution against using measurements of the size of the emission region to derive values such as the gamma of the blast wave/ejecta/shocked material, except in special circumstances. Relatedly, we note that the term spreading can be confusing. A shell can spread laterally (diverge or expand perpendicular to its propagation), or it can spread radially (rarefy or expand along the direction of propagation). In this work, we only ever use the term to mean {\em lateral spreading} (see section~\ref{sec:discuss_lateral}).

\section{Critical distances}
\label{sec:critical}

\begin{figure*}
\centering
\includegraphics[width=\linewidth]{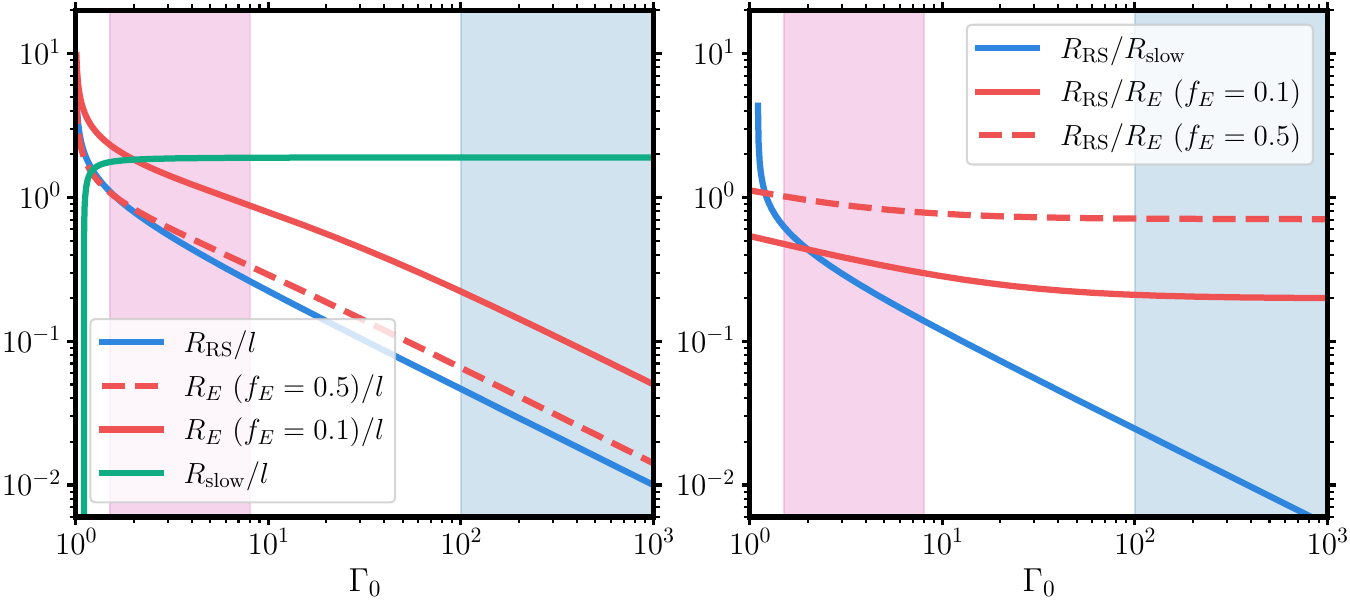}
\caption{Comparison of the various analytic critical distances/radii defined in the text. The left hand panel shows the various length scales normalised to the Sedov-Taylor-von Neumann length ($l$): the reverse shock crossing distance, $R_{\rm RS}$, the slowing distance, $R_{\rm slow}$, and the energy dissipation scale, $R_E$ for two values of $f_E$ (the fraction of the initial energy that has been dissipated). The right-hand panel shows ratios of $R_{\rm RS}$ to the other critical radii. Reverse shocks are short-lived in GRBs but long-lived in XRBs. XRBs dissipate their energy on around the same scale over which they slow down, whereas GRBs dissipate most of their energy before they have reached a low Lorentz factor. The pink and blue shaded regions show approximate initial Lorentz factor ranges for XRBs and GRBs.}
\label{fig:critical_radii}
\end{figure*}

Following, e.g., \cite{sari_piran_1995}, we define a series of critical radii, each of which have an associated observer frame timescale (see section~\ref{sec:reverse_shock_time}). The critical radii represent the value of $R$, the laboratory frame distance from the origin to the blast wave outer edge/shock (or to the ejecta/shell), at a given critical point in the evolution.These are:
\begin{itemize}
    \item $R_{\rm RS}\equiv R(t_{\rm RS})$: the point at which the reverse shock finishes crossing the ejecta (this is $R_\Delta$ in the notation of \citealt{sari_piran_1995}).
    \item $R_\gamma\equiv R(t_\gamma)$: the point at which the swept up mass equals $1/\Gamma_0$ of the initial mass (sometimes called the deceleration radius in the GRB community)
    \item $l \equiv R(t_l)$: the angle-corrected Sedov-Taylor-von Neumann length scale; the point at which the rest mass energy of the swept up material equals the initial energy 
    \item $R_{\rm slow} \equiv R(t_{\rm slow})$: the point at which the blast wave slows to a low Lorentz factor $\Gamma_{\rm slow}$
    \item $R_{\rm E} \equiv R(t_{\rm E})$: the energy dissipation radius, the point at which the blast wave has dissipated a fraction $f_E$ of its energy 
\end{itemize}
These critical radii can be defined in general terms, or specifically for a kinetically-dominated initial condition with $E_0 = (\Gamma_0-1) M_0 c^2$. We now define or derive expressions for each of these based on mass and energy conservation; the first three are also given by \cite{sari_piran_1995} who discuss them in the GRB (i.e. on axis, ultra-relativistic) context. The critical radii are independent of observer viewing angle, but the critical times do depend on viewing angle, from equation~\ref{eq:drdt}, as discussed in the next section. Each critical radius is given for $k=0$ (where $\rho(R) \propto R^{-k}$), but can be generalised to $k \neq 0$ by replacing the $1/3$ exponent with $1/(3-k)$. We plot $R_{\rm RS}$, $R_{\rm slow}$ and $R_E$ in Fig.~\ref{fig:critical_radii}, and now derive and discuss them in turn. In the equations, we assume $\Gamma_{\rm sh} \approx \Gamma$, but in the plots we make use of the more accurate equation~\ref{eq:gammash}.

\subsection{Reverse shock crossing}
There are various subtleties in choosing the appropriate expression for the reverse shock crossing radius; it depends somewhat on the shock and shell characteristics, in particular whether or not the shell is thin or thick and whether the reverse shock is relativistic or Newtonian \citep{sari_piran_1995,kobayashi2000,kobayashi_numerical_2000}. Going forward we assume equivalence between $R_{\rm RS}$ and $R_{\rm \gamma}$, which is appropriate for the thin shell Newtonian regime and was the behaviour in our numerical hydrodynamic simulations in the moderately relativistic regime (see Appendix~\ref{app:hydro}). This assumption is equivalent to saying that the reverse shock crosses the shell when the swept up mass equals $1/\Gamma_0$ of the shell mass, which happens when $R$ satisfies
\begin{equation}
\frac{\Omega \rho_0 R^3}{3} = \frac{M_0}{\Gamma_0} \, . 
\end{equation}
The reverse shock should then cross the shell at the critical radius 
\begin{equation}
R_{\rm RS} = 
\left (\frac{3 M_0}{\Omega \rho_0 \Gamma_0} \right)^{1/3} = 
\left (\frac{3 E_0}{\Omega \rho_0 c^2 \Gamma_0 (\Gamma_0 - 1)} \right)^{1/3} \, .
\label{eq:r_rs}
\end{equation}
This equation tells us that, for the same energy input, more relativistic shells have their RS crossing earlier on in the evolution, and the opposite is true for more massive shells. Similarly, for constant mass, shells that are more relativistic also have their RS crossing earlier on in the evolution, although the effect is less pronounced ($\propto \Gamma_0^{-1}$) than for constant energy. The reason for this behaviour is relatively straightforward. The Lorentz factor is the ratio of kinetic energy to rest mass energy and this critical radius is defined by mass sweeping; lower Lorentz factor shells have higher masses for a given energy, so hit the critical shock crossing point later in their evolution.  

\subsection{The relationship between $R_{\rm RS}$ and $l$}
The slowing down of the ejecta is determined by energy conservation, so we might expect significant slowing when the rest mass energy of the swept up material equals the initial energy. For a kinetically dominated flow, this gives
\begin{equation}
\frac{\Omega \rho_0 c^2 R^3}{3} = (\Gamma_0 - 1) M_0 c^2. 
\end{equation}
Thus again we have a critical radius which is equal to the Sedov-Taylor-von Neumann length given previously
\begin{equation}
l = \left(\frac{3(\Gamma_0 - 1) M_0}{\Omega \rho_0} \right)^{1/3} \, ,
\end{equation}
which is also often described \citep[e.g.][]{piran_gamma-ray_1999} as the point at which the blast wave becomes non-relativistic or Newtonian (although see sections~\ref{sec:slowing} and \ref{sec:evolution}).
Taking the ratio of $R_{\rm RS}$ and $l$ is instructive:
\begin{equation}
\frac{R_{\rm RS}}{l} = \left[\Gamma_0(\Gamma_0 - 1)  \right]^{-1/3}.
\label{eq:lorentz_ratio}
\end{equation}
This quantity is plotted in the left-hand panel of Fig.~\ref{fig:critical_radii}. We find the ratio $R_{\rm RS}/l$ only depends on how relativistic the flow is. This result is similar to that obtained in GRBs \citep[see, for example][]{sari_piran_1995,piran1999} where the ultra-relativistic case gives $R_{\rm RS} \approx \Gamma_0^{-2/3} l$. For these highly relativistic flows, the reverse shock is relatively short lived: $R_{\rm RS} \ll l$. For the moderately relativistic or transrelativistic case, applicable to XRBs and some other scenarios, we expect $l / R_{\rm RS} \sim {\rm a~few}$. Equation~\ref{eq:lorentz_ratio} implies that measurement of $l$ and $R_{\rm RS}$ or equivalent timescales in XRBs can be used to constrain the initial Lorentz factor, a possibility that we discuss further in section~\ref{sec:lorentz_discuss}.

\subsection{Slowing}
\label{sec:slowing}
\cite{zdziarski_cause_2024} define a slowing distance -- which is more accurate at characterising the slowing of the ejecta, for modest $\Gamma_0$, than $l$ -- as the distance at which the blast wave has slowed to a Lorentz factor of ${\cal K}\Gamma_0$ (where ${\cal K}$ is a positive constant less than one). This distance is calculated from the transfer of energy to the shocked ISM through equation~\ref{eq:energy_cons}, rather than comparing to the rest mass energy swept up. They give the equation
\begin{equation}
R_{\cal K}\approx \left[
\frac{3 (1-{\cal K})\Gamma_0 E_0}
{\Omega \rho_0 c^2 \sigma (\Gamma_0 - 1) ({\cal K}^2 \Gamma_0^2 - 1)} 
\right]^{1/3} \, ,
\end{equation}
which can be derived from equation~\ref{eq:energy_cons} if one sets $\Gamma={\cal K} \Gamma_0$ and approximates $\Gamma_{\rm sh} \approx {\cal K} \Gamma_0$. However, since this equation involves a fractional decrease in $\Gamma$, it is not suitable for comparing the stopping distance of blast waves in vastly different $\Gamma_0$ regimes. Instead we calculate the distance $R_{\rm slow}$ at which the blast wave has reached a given low Lorentz factor $\Gamma_{\rm slow}<\Gamma_0$, which happens when 
\begin{equation}
E_0 = \frac{\Gamma_{\rm slow} - 1}{\Gamma_0 - 1} E_0 - \frac{\Omega \sigma \rho_0 c^2 R_{\rm slow}^3}{3} (\Gamma_{\rm sh}^2 - 1).
\end{equation}
We can then rearrange for $R_{\rm slow}$ and approximate $\Gamma_{\rm sh} \approx \Gamma_{\rm slow}$, obtaining
\begin{equation}
R_{\rm slow} \approx \left[
\frac{3 (\Gamma_0-\Gamma_{\rm slow}) E_0}
{\Omega \rho_0 c^2 \sigma (\Gamma_0 - 1) (\Gamma_{\rm slow}^2 - 1)} 
\right]^{1/3} \, .
\label{eq:rslow}
\end{equation}
We adopt $\Gamma_{\rm slow} = \sqrt{5}/2 \approx 1.118$ for our calculations, corresponding to $\beta \Gamma = 1/2$. 

We plot $R_{\rm slow}/l$ in the left-hand panel of Fig.~\ref{fig:critical_radii}. From this figure and equation~\ref{eq:rslow}, $\Gamma_0 \gg \Gamma_{\rm slow}$ one finds that the deceleration scale defined this way is, for fixed $E_0$, {\em independent of $\Gamma_0$}; the figure of merit for determining the stopping or slowing distance is then $[E_0/(\rho_0 \Omega)]^{1/3}$ as expected from the generalized Sedov-Taylor-von Neumann scale. Ultra-relativistic and moderately relativistic blast waves thus decelerate to a given low $\Gamma$ at the same radius for a given energy, as long as they are embedded in the same density environment and have the same opening angle. The difference is that moderately relativistic blast waves have a longer-lived reverse shock phase, and have used up much less of their energy by the time they decelerate to a low $\Gamma$. The reverse shock and Sedov-Taylor-von Neumann phases are therefore comparatively more important in moderately relativistic blast waves than ultra-relativistic ones. 

\subsection{Energy dissipation}
Finally, we derive the energy dissipation radius: the radius at which the ejection has only a factor $f_E$ of its initial energy. This happens when 
\begin{equation}
E_0 = f_E E_0 + \sigma (\Gamma_{\rm sh}^2 - 1) m_{\rm sw} c^2 
\end{equation}
since $\Gamma-1 = E/(M_0 c^2)$ one can derive that $\Gamma=f_E (\Gamma_0-1) + 1$ which, under the assumption $\Gamma_{\rm sh} \approx \Gamma$ gives 
\begin{equation}
E_0 = f_E E_0 + \sigma m_{\rm sw}(R_E) c^2 \left[ 
f_E^2 (\Gamma_0-1)^2 + 2f_E (\Gamma_0 - 1)
\right]
\end{equation}
which after some manipulation gives
\begin{equation}
R_E = \left[
\frac{3 (1-f_E) E_0}
{\Omega \rho_0 c^2 \sigma (f_E^2 (\Gamma_0-1)^2 + 2f_E (\Gamma_0 - 1))} 
\right]^{1/3}.
\end{equation}

The left-hand panel of Fig.~\ref{fig:critical_radii} shows each of these critical radii, normalised to $l$, as a function of $\Gamma_0$, for two choices of $f_E=0.1,0.5$. The right hand panel then shows the ratios $R_{\rm RS}/R_{\rm slow}$ and $R_{\rm RS}/R_E$, the latter for the same two values of $f_E$. For fixed $E_0$, $l$ is independent of $\Gamma_0$, and $R_{\rm slow}$ becomes independent of $\Gamma_0$ once $\Gamma_0 \gg \Gamma_{\rm slow}$. In the XRB regime ($\Gamma_0 \sim {\rm a~few}$), $R_{\rm RS}$ is less than but comparable to $R_{\rm slow}$, with $R_{\rm RS}/R_{\rm slow}$ ranging from $\sim 0.1-0.5$. However, in the ultra-relativistic regime $R_{\rm RS}/R_{\rm slow} \sim 0.01$, and the reverse shock crosses the shell at small $R$, i.e. when the blast wave is still very compact.

\section{Timescales and Evolution}
\label{sec:timescales}
\subsection{Reverse shock timescales}
\label{sec:reverse_shock_time}
We also seek to calculate the critical time (in the observer frame) associated with the reverse shock crossing, since this dictates the window during which an observer will be able to search for signatures of the reverse shock. For the on-axis, ultra-relativistic case, the time can be estimated straightforwardly from integrating equation~\ref{eq:drdt_grb}, but for the more general case we must instead integrate equation~\ref{eq:drdt_xrb} until the critical radius in question is reached. We do this by numerical solution for a range of $\Gamma_0$, using the procedure described by \cite{Carotenuto_2022}, and calculate $t_{\rm RS}$. In the next subsection, we also calculate $t_{\rm slow}$ using the same procedure. 

\begin{figure}
\centering
\includegraphics[width=\linewidth]{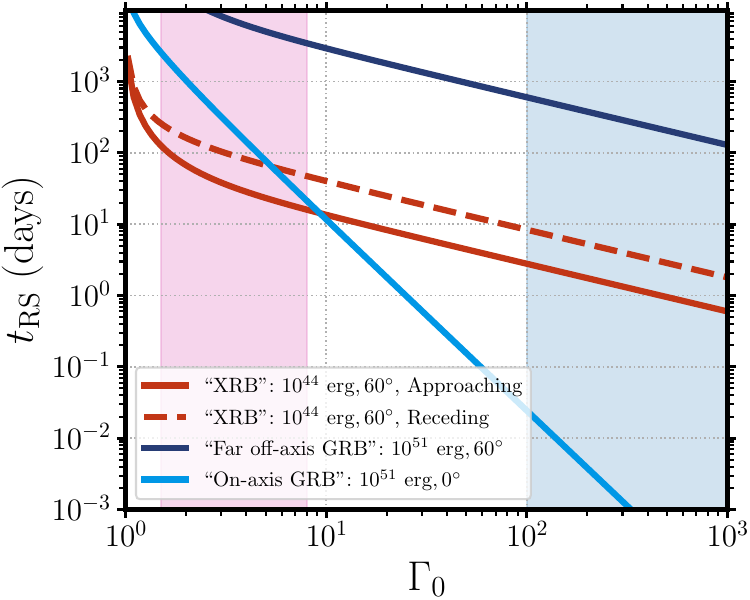}
\caption{Reverse shocks live longer when blast waves are less relativistic and/or more off-axis. Reverse shock timescales, calculated from numerical integration of the external shock model, as a function of $\Gamma_0$ for different $E_0$ and $\theta$, corresponding approximately to an off-axis XRB ($E_0=10^{45}~{\rm erg~s}^{-1}, \theta=60^\circ$), an, off-axis GRB ($E_0=10^{51}~{\rm erg~s}^{-1}, \theta=60^\circ$) and an on-axis GRB ($E_0=10^{51}~{\rm erg~s}^{-1}, \theta=0^\circ$). For the XRB, we show $t_{\rm RS}$ for both the approaching and receding components. The pink and blue shaded regions show approximate initial Lorentz factor ranges for XRBs and GRBs.}
\label{fig:times}
\end{figure}

Fig.~\ref{fig:times} shows the reverse shock crossing time, $t_{\rm RS}$, defined such that $R(t_{\rm RS})\equiv R_{\rm RS}$, as a function of $\Gamma_0$ for four choices of energies and viewing angles. These choices correspond approximately to an off-axis XRB ($E_0=10^{45}~{\rm erg~s}^{-1}, \theta=60^\circ$; approaching and receding components), an off-axis GRB ($E_0=10^{51}~{\rm erg~s}^{-1}, \theta=60^\circ$) and an on-axis GRB ($E_0=10^{51}~{\rm erg~s}^{-1}, \theta=0^\circ$). The figure shows that reverse shock timescales in XRBs are typically tens of days if $\Gamma_0 \lesssim 5$, whereas in an on-axis GRB $t_{\rm RS} \lesssim 0.1~{\rm day}$. By comparing the ``Far off-axis GRB'' and ``XRB'' cases it becomes clear that for more energetic ejecta, the reverse shock timescale is longer, by $t_{\rm RS} \propto E_0^{1/3}$ for fixed $\Gamma_0$ and $\theta$. This figure emphasizes key aspects of moderately and off-axis relativistic systems: their reverse shocks are, in principle, observable for much longer than their ultra-relativistic, on-axis cousins.  

The observability of reverse shock signatures is not only determined by the reverse shock crossing time. In GRBs, the peak in the reverse shock component of the light curve typically occurs at $<1$~day, but does not usually correspond to the point of reverse shock crossing. In the majority of events where the reverse shock has been identified, it has been through the discovery of a fast declining early-time component \citep{laskar2013,laskar2016,laskar2019}. To our knowledge, there is only one event where the peak (the synchrotron self-absorption break) of the reverse shock has been track through multiple observing bands \citep{bright_rhodes_2023,rhodes2024}. The detection of the emission from reverse shock electrons can then continue for a number of days post-shock crossing, as the particles accelerated by the reverse shock gradually cool due adiabatic losses as the shocked plasma expands, synchrotron losses, and inverse Compton losses.

Moderately relativistic XRB ejecta resolved from the core are observed as optically thin throughout their evolution. This, coupled with the fact that the reverse shock crossing occurs on 10-100 day timescales, means that the peak of the reverse shock emission will generally correspond to the completion of the reverse shock crossing, as also found by \citesavard. The relevant timescale over which a reverse shock signature can be observed is then approximately $t_{\rm RS, obs} \approx t_{\rm RS} + {\rm min} (\tau_{\rm ad}, \tau_{\rm rad})$. Here, $\tau_{\rm ad}$ is the observer frame adiabatic cooling timescale, and $\tau_{\rm rad}$ is the frequency dependent radiative observer frame cooling time, including synchrotron and inverse-Compton losses. In reality, both of these timescales will evolve, and vary from source to source, since they depend on the local plasma conditions. We have investigated the details of the RS crossing and its light curve in a forthcoming companion paper \citep{cooper_reverse}.

\subsubsection{The receding component}
\label{sec:receding}
The receding ejecta component or shell is not detected in GRBs; their on-axis, ultra-relativistic nature means the radiation is invariably strongly beamed away from the observer (although see \citealt{granot2018,fernandez2022} for predicted counter-jet effects on the GRB radio centroid, and \citealt{li2024} for multiwavelength counterjet prospects in GW170817). However, in XRBs it is relatively common to detect the receding component, particularly in far off-axis (i.e. high inclination) sources such as MAXI J1820$+$070 \citep{bright_extremely_2020}, XTE J1550–564 \citep{corbel_large-scale_2002} and MAXI J1848-015 \citep{bahramian2023}; see also discussion by \cite{maccarone2022}. The behaviour of $dR/dt$ or $\alpha(t)$ for receding ejecta can be obtained by taking the alternative sign in the denominator of equations~\ref{eq:drdt_xrb} and \ref{eq:alpha_xrb}, so it is straightforward to carry out the same calculation of $t_{\rm RS}$ in the receding case. In Fig.~\ref{fig:times}, we show this RS timescale for the receding jet with a dashed line. As expected, the timescale is longer in the receding jet case, by the ratio of the integral of the Doppler factors over the evolution from launch to the point at which $R=R_{\rm RS}$. This is of course generically true for any observer-frame critical timescale. In the highly relativistic limit, the ratio simplifies, so that the RS timescale is longer for the receding ejecta compared to the approaching ejecta by a factor  $(1+\cos\theta)/(1-\cos\theta)$, which is $3$ for the plotted case of $\cos \theta = 1/2$. Reverse shock signatures will therefore be observable for even longer in XRBs where the receding jet is detectable. 

\subsection{Distinguishing the reverse and forward shocks}
As well as operating on different timescales, electrons accelerated by the reverse and forward shocks reside in spatially distinct regions. To investigate this, we ran a 1D relativistic hydrodynamic simulation of a spherical shell with $\Gamma_0 = 2.5$ and $E_{k,{\rm iso}} = 1.313 \times 10^{48}~{\rm erg}$. The numerical method and setup is described in appendix~\ref{app:hydro}. We identified the point of reverse shock crossing and also tracked the location of the forward shock and ejecta material, the latter being a proxy for what would be observed if reverse shock emission was powering what we observed in XRB transient ejecta. The time of evolution of the radius $R$ of these two locations, normalised to $l$, is shown in Fig.~\ref{fig:shock_radii}. At early times (where $R<R_{\rm RS})$, the two shocks move in concert and the separation between shocked plasma is small. However, at later times, after the reverse shock crossing, the forward shock decouples from any reverse shocked material and propagate ahead of any electrons accelerated by the reverse shock. {An important consequence of this behaviour is that, if the emission comes mainly from reverse shocked electrons, the angular separation may increase more slowly for the same total energy input $E_0$; equivalently, this could mean that energy estimates obtained with a forward shock kinematic model could be conservative underestimates.

XRB jet ejections can be spatially resolved, in terms of their separation from the core and sometimes also in terms of the structure of the ejecta itself. Thus, in these sources the spatial decoupling of the forward and reverse shocks provides another way to search for signatures of different shock components (in addition to the aforementioned combined kinematic and radiative modelling). In particular, it may be possible to observe stratified emission structures in radio blobs as a result of the distinct shocked regions, or multiple components diverging from each other in core separation over time. We note that MAXI J1848$-$015 does have radially extended structure in VLA images \citep{bahramian2023}, and its large-scale jets are likely to have a particularly long-lived reverse shock given the high inclination and relatively modest velocity of $\approx 0.8 c ~(\Gamma\approx 1.67)$. In addition, changes in relative brightness of the forward and reverse shocked regions could lead to jittering or an illusion of fast changes in apparent angular separation. There are hints of both of these effects in some XRB sources, but it is challenging to know if multiple shocks are responsible. However, in general, the spatially resolved nature of jet ejections in XRBs provides a powerful pathway to understand astrophysical reverse shocks. 

One final way to distinguish between reverse and forward shock signatures is via radio polarization measurements. The reverse and forward post-shock regions could have quite different magnetic field topologies and densities of cold electrons, which could translate to a change in polarization properties if the synchrotron emission changes from reverse to forward shock dominated. This could manifest itself as a rotation in the polarization angle due to different field geometries, an evolution in fractional linear polarization due to the different degrees of ordering of the magnetic field, or changes in internal Faraday rotation measure. Related to this possibility, \cite{laskar2019} argue that the linear polarization in the the millimetre-band afterglow of GRB 190114C can be attributed to a reverse shock. In addition, from spatially resolved polarization data, \cite{orienti2017} find evidence for disordered magnetic fields in the hotspot of radio galaxy 3C 445. Both of these studies used data from the Atacama Large Millimeter Array, and similar approaches in either the mm or radio bands could prove fruitful in XRBs.

\begin{figure}
\centering
\includegraphics[width=\linewidth]{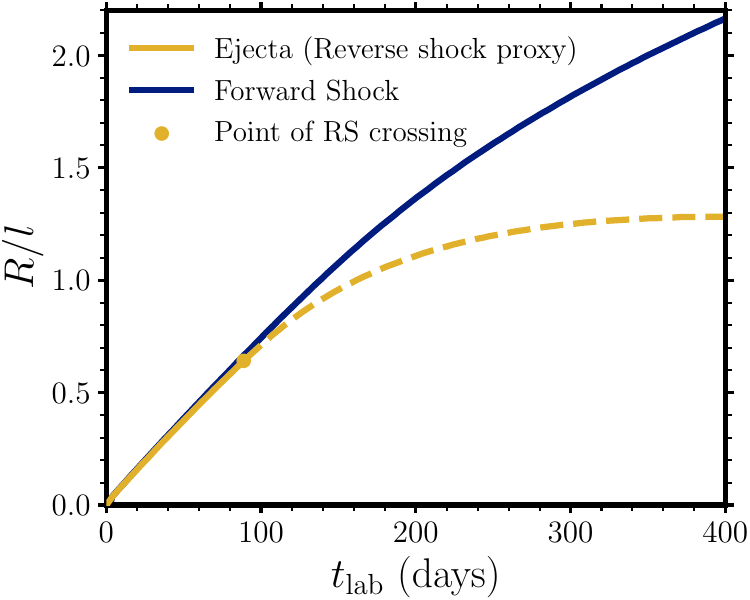}
\caption{Comparison of the hydrodynamic evolution of the forward shock position and the maximum distance of shell material/ejecta (a proxy for the `reverse shock signature'). Both curves are shown as a function of laboratory frame time, $t_{\rm lab}$, equivalent to our observer frame time at $\theta_i=90^\circ$. The circle marks the point of reverse shock crossing, at which the ejecta material detaches from the forward shock. The results are obtained from a 1D relativistic hydrodynamic simulation with $\Gamma_0=2.5$.}
\label{fig:shock_radii}
\end{figure}

\subsection{Overall Evolution}
\label{sec:evolution}

\begin{figure*}
\centering
\includegraphics[width=\linewidth]{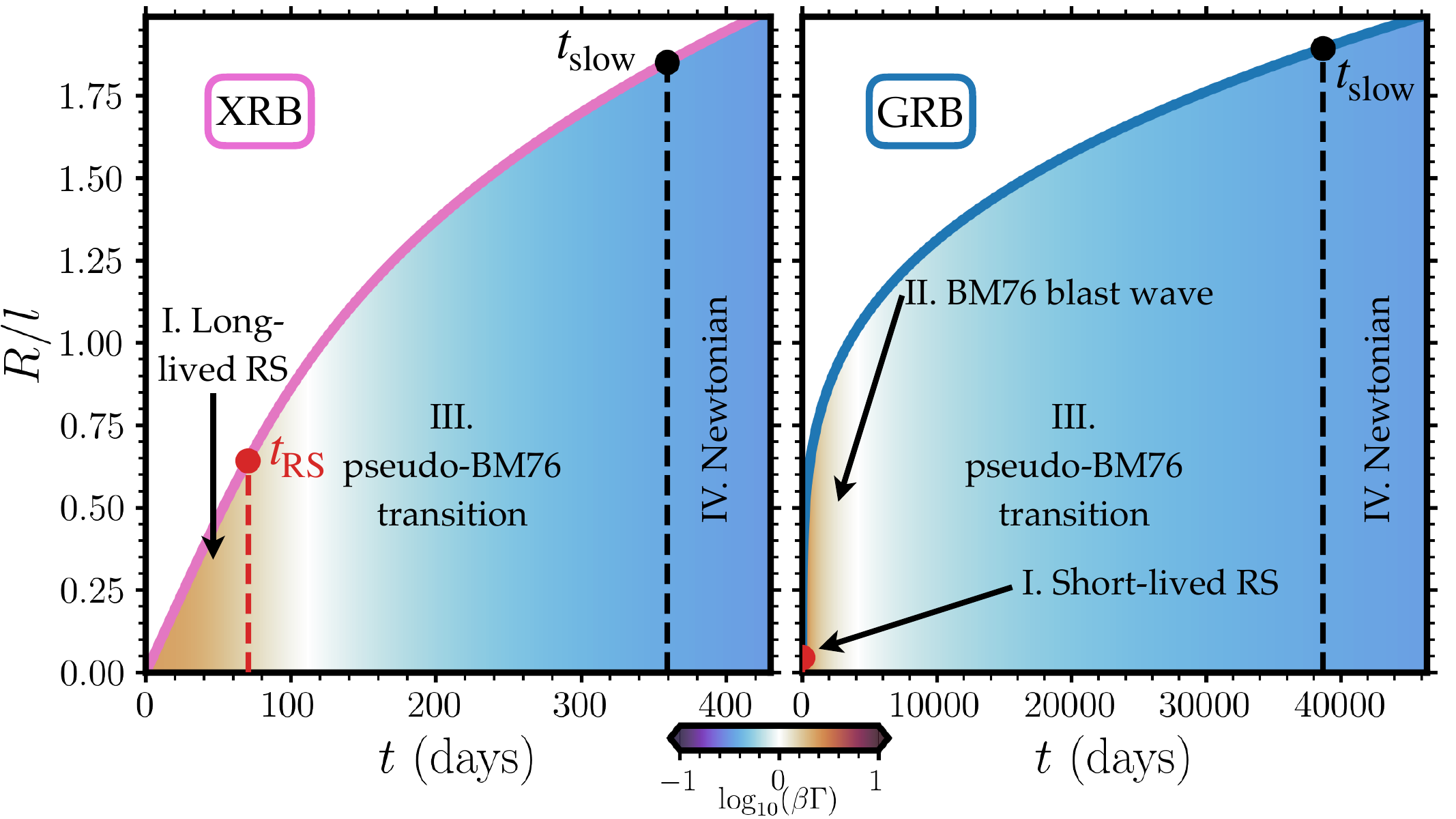}
\caption{Schematic figure showing the evolution of $R$, the distance from the central engine to the blast wave shock, over observer-frame time in approximate XRB ($E_0=10^{44}~{\rm erg~s}^{-1}$, $\theta_i=60^\circ$, $\Gamma_0=2.5$; left panel) and GRB ($E_0=10^{51}~{\rm erg~s}^{-1}$, $\theta_i\approx0^\circ$, $\Gamma_0=100$; right panel) regimes, both for the approaching component. The curve is normalised to the Sedov-Taylor-von Neumann length, $l$. In both cases, the $R(t)$ curve has been obtained from numerical integration of equations~\ref{eq:energy_cons} and \ref{eq:drdt}, and two of the critical radii are marked, as are the important phases in the evolution. The colourmap in the background denotes $\log \beta \Gamma$. The parameter values are as defined in section~\ref{sec:represent}, with $n_0=6.95 \times 10^{-3}~{\rm cm}^{-3}$ and $\phi = 1^\circ$. A version of this figure on logarithmic axes is shown in Fig.~\ref{fig:schematic_log}.  
}
\label{fig:schematic}
\end{figure*}

In Fig.~\ref{fig:schematic}, we show a schematic diagram depicting the evolution of $R(t)$ with two of the critical timescales ($t_{\rm RS}$ and $t_{\rm slow}$) and phases of different behaviour marked and labeled I-IV. In both cases, the $R(t)$ curve has been obtained from numerical integration of equations~\ref{eq:energy_cons} and \ref{eq:drdt}, and the curve is normalised to $l$ making the magnitude of $R$ independent of $E_0$ (although not of $\Gamma_0$ or $\theta$). A version of the schematic figure with logarithmic axes is given in the appendix (Fig.~\ref{fig:schematic_log}), in which it is easier to see the dynamic range of times and scales as well as the various limiting slopes of the $R(t)$ curves. We now, in turn, briefly discuss the qualitative evolution of the GRB and XRB scenarios depicted. We also refer the reader to \cite{peer2012} who uses a similar shell/external shock model to plot the evolution of $\Gamma$, $\beta \Gamma$ and $R$ over observer time in the XRB and GRB regimes, for somewhat different parameters. 

The evolution of ultra-relativistic blast waves has been extensively discussed in the literature \citep[e.g.][]{sari_piran_1995,sari1997,piran1999} and so we only briefly review the key phases here. After, or sometimes during, the prompt emission phase, there is a short-lived reverse shock crossing phase. The behaviour of the shock depends on the relative timescales of the shock crossing and the burst duration, which determines whether or not the thin or thick shell regime is relevant and also whether the reverse shock is relativistic or Newtonian \citep{sari_piran_1995}. After the reverse shock crossing, the blast wave is still highly relativistic and the system enters the Blandford-Mckee phase (phase II). This phase is long-lived and results in the well known $\Gamma \propto t^{3/8}$, $R \propto t^{1/4}$ evolution. Eventually, at around $t_l$ or $t_{\rm slow}$, the system gradually transitions, via a mildly relativistic pseudo-Blandford-Mckee phase (III), to a truly Newtonian, Sedov-Taylor-von Neumann solution (phase IV). The recovery of the true Sedov-Taylor-von Neumann evolution can take some time; Fig~\ref{fig:schematic_log} shows that even at $R/l \sim 3$ the gradient of the $R(t)$ curve has steepened, but has not yet reached the limiting $R\propto t^{2/5}$ behaviour. This result is consistent with \cite{zhang2009}, who find that the evolution can be described by the Sedov-Taylor-von Neumann solution after $R/l \sim 5$ \citep[see also][for further discussion of the transition to the non-relativistic regime]{waxman1998,livio2000,peer2012,jetsimpy2024}.

In the moderately relativistic case, the reverse shock phase (I) lasts much longer, as previously discussed. At the time the reverse shock finishes passing through the shell/ejecta the blast wave is not highly relavistic as in the GRB case, and thus the system never enters a true Blandford-Mckee phase. However, the system is also not strictly ``Newtonian'', at least  in the sense that $\beta \Gamma \gtrsim 1$ and the system has not started following the expected Sedov-Taylor-von Neumann evolution. We thus refer to this phase III as a pseudo-Blandford-Mckee phase  where the system is not described accurately by either of the self-similar solutions used to model blast waves. This pseudo-Blandford-Mckee phase III is also the phase in which lateral spreading is likely to be important (see section~\ref{sec:discuss_lateral}). Finally, the system eventually transitions to a Sedov-Taylor-von Neumann phase (IV) where $R\propto t^{2/5}$ and the expansion is non-relativistic. 

\section{Geometry}
\label{sec:geometry}

\subsection{Cloud crushing and arbitrary geometries}
\label{sec:discuss_cloud}
\citesavard\ model an XRB jet ejection as a propagating moderately relativistic cloud or blob, which, in the frame of the cloud, is a relativistic analogue of the well-studied cloud-crushing problem \citep{klein_hydrodynamic_1994}. A reverse shock passes through the cloud, and a forward shock is driven into the surrounding medium. The cloud disrupts on a few cloud-crushing timescales.  The behaviour is qualitatively similar to that considered here, and in fact \citesavard\ also find that 1D simulations with the same properties produce very similar deceleration profiles to the 2D cloud geometry. The 2D cloud geometry is somewhat similar to the boosted fireball model considered by \cite{duffell2013}. 

The main differences between the relativistic cloud-crushing picture and the quasi-spherical external shock model are purely geometric. This can be seen by considering the swept up mass, $m_{\rm sw}$ from equation~\ref{eq:mass_swept}, which is only appropriate for a quasi-spherical outflow. The most general form of $m_{\rm sw}$ comes from considering an infinitesimal mass element $dm = \rho A dr$, where $A$ is the cross-sectional area of the element and both $A$ and $\rho$ are, in general, functions of $R$. This gives 
\begin{equation}
    m_{\rm sw}(R) = \int^R_0 \rho(R^\prime) A(R^\prime) dR^\prime.
    \label{eq:mass_general}
\end{equation}
which can be used for {\em any} well-defined 3D shape swept out in the ISM by the blast wave. If we assume a blob of constant radius, $r_b$, and constant density $\rho$ then the mass swept up at a radius $R$ is just set by the volume of a cylinder, that is  
\begin{equation}
    m_{\rm sw}(R) = \pi r_c^2 R \rho_0. 
\end{equation}
Following the same procedure as in section~\ref{sec:critical}, for constant cloud radius $r_b$ we can find the radius at which the swept up mass equals $M_0/\Gamma_0$, which gives
\begin{equation}
    R_{\gamma, c} = \frac{4 r_c}{3 \Gamma_0} \chi
\end{equation}
where $\chi=\rho_c/\rho$ is the density contrast and we have used $M_0=(4/3) \pi r_c^3 \rho_c$ for the initial cloud mass. This equation can also be written in terms of energy as 
\begin{equation}
    R_{\gamma, c} = \frac{E_0}{\pi r_c^2 \Gamma_0 ( \Gamma-1) \rho_0 c^2}.
\end{equation}

There are two interesting scalings in the above equations that are worth discussing. The first is that $R_{\gamma, c} \propto \chi r_c$, which, in the non-relativistic limit, matches the scaling of the drag timescale, $t_{\rm drag}$, quoted by \cite{klein_hydrodynamic_1994}; $t_{\rm drag}$ is the characteristic timescale over which the cloud decelerates due to drag forces. However, it is in contrast to the canonical cloud-crushing time $\chi$ dependence of $t_{\rm cc} \propto \chi^{1/2} r_c$. This difference results in $t_{\rm cc} < t_{\rm drag}$, and the conceptual difference between the two timescales is the same as that between $t_\gamma$ and $t_{\rm RS}$ in the relativistic blast wave case \citep[][see also section~\ref{sec:critical}]{sari_piran_1995}. However, in a proper hydrodynamic model, the idealised picture breaks down and $t_{\rm cc}$ 
is rather similar to $t_{\rm drag}$ \citep{klein_hydrodynamic_1994}; both should be thought of as merely characteristic timescales. 
The second important scaling is that $R_{\gamma, c} \propto E$, in contrast to the $R_{\rm RS} \propto E^{1/3}$ scaling in the quasi-spherical or conical case. This different scaling arises purely from the linear dependence of $m_{\rm sw}$ on $R$, and is unlikely to be correct. As noted by \cite{klein_hydrodynamic_1994} and \citesavard, the cylindrical approximation for the swept up mass is an oversimplification; in reality, the cloud expands and distorts, and the true form of $m_{\rm sw}(R)$ will be more complex. In principle, the general form of the swept up mass (equation~\ref{eq:mass_general}) can be implemented within the numerical solution of equation~\ref{eq:energy_cons} to obtain, e.g., $R(t)$ for arbitrary geometries.

\subsection{Lateral spreading}
\label{sec:discuss_lateral}
The process of lateral expansion, often termed ``spreading'', by which we mean an opening angle increasing over time and as a function of $R$ ($d\phi/dR > 0$), is often discussed in a GRB context and there have been a number of detailed analytic and numerical studies \citep[e.g.][]{rhoads1999,livio2000,zhang2009,Lyutikov2012,vaneerten2012,granot2012,granot2018,duffell2018,fernandez2022,Govreen-Segal2024}. Early on in GRB evolution, the different regions in the jet are not causally connected, and no spreading occurs. However, spreading is thought to start when  $\Gamma \phi \lesssim 1$, because at that point the edges of the jet are causally connected. Recently, \cite{Govreen-Segal2024} used both numerical and analytic methods to show that, in this spreading phase, the Lorentz factor in the wings follows $\Gamma(\varphi) \propto \varphi^{-1.5}$ where $\varphi$ is the angle from the axis. Lateral spreading effects can also be approximately accounted for using modelling tools such as \texttt{afterglowpy} \citep{ryan_afterglowpy2020} or \texttt{jetsimpy} \citep{jetsimpy2024}. \cite{Govreen-Segal2024} discuss a range of observational consequences induced by this lateral spreading including the impact on the light curve rise and spectral evolution. However, our principal interests here are how, in XRBs, lateral spreading might affect the separation over time -- i.e. $dR/dt$ or $\alpha(t)$ -- or the shock properties.

\begin{figure}
\centering
\includegraphics[width=\columnwidth]{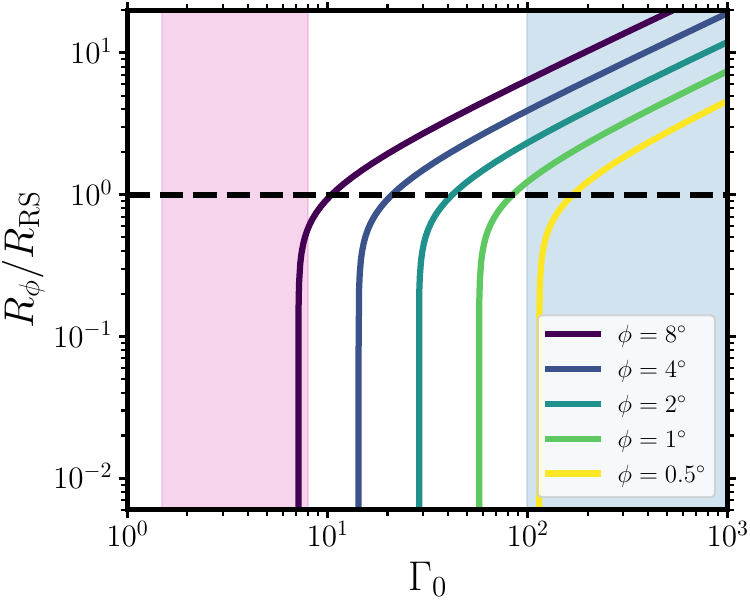}
\caption{The onset of lateral spreading is set by RS crossing in XRBs and mildly relativistic blast waves. The solid lines, colour-coded by half-opening angle $\phi$, show the ratio of $R_{\rm \phi}$, the radius at which $\Gamma \phi = 1$, to $R_{\rm RS}$, the reverse shock crossing radius as a function of $\Gamma_0$. For $\Gamma_0 < 1/\phi$ the equation becomes meaningless. $R_{\rm \phi}$ is estimated analytically from equation~\ref{eq:r_phi}.
Above the dotted line (and at high $\Gamma_0$), the critical Lorentz factor for spreading is reached after the RS crossing phase. Below the dotted line, the onset of lateral spreading is instead determined by the point of RS crossing. 
}
\label{fig:lateral1}
\end{figure}

Based on the above $\Gamma \phi \lesssim 1$ criterion, we can define a critical Lorentz factor below which lateral spreading is important as $\Gamma_{\rm spread} = (180/\pi) \phi_\circ^{-1} \approx 57 \phi_\circ^{-1}$, where $\phi_\circ$ is now the opening angle in degrees. XRB jet ejections have inferred opening angles of only up to a few degrees \citep{Miller-Jones2006}. Thus, taken at face value, this equation indicates that lateral spreading is likely to be important in XRB jet ejections. However, there is a big difference between the GRB and XRB case: in the former, the RS has always crossed the ejecta before $\Gamma \phi \lesssim 1$, whereas in the latter case this is not true. To see this, we define a critical radius $R_\phi$, the value of $R$ at which $\Gamma \phi = 1$. This radius can be derived in a similar manner to the critical radii in the previous section, and is given by 
\begin{equation}
    R_\phi = \left[
\frac{3 (\Gamma_0-\Gamma_{\rm spread}) E_0}
{\Omega \rho_0 c^2 \sigma (\Gamma_0 - 1) (\Gamma_{\rm spread}^2 - 1)} 
\right]^{1/3} \, ,
\label{eq:r_phi}
\end{equation}
which is only valid for $\Gamma_0 > \Gamma_{\rm spread}$. In Fig.~\ref{fig:lateral1} we show the value of $R_\phi/R_{\rm RS}$ as a function of $\Gamma_0$, for various values of $\phi$. We see that $R_\phi<R_{\rm RS}$ only at relatively low $\Gamma_0$, as one would expect. In XRBs, we will almost always be in the regime where $\Gamma_0 > \Gamma_{\rm spread}$. 

During the RS crossing phase, the shell {\em does not spread laterally}. This is because the shell is cold and has only radial momentum. The material that has passed through the post-shock is hot, and can spread sideways, but will flow away from the RS region and not impede the shell's progress significantly. The kinetically dominated shell thus continues to drive the blast wave forwards and the overall behaviour will depend somewhat on the detailed hydrodynamics and density contrast of the shell. Thus, from Fig.~\ref{fig:lateral1} and the above argument, in moderately/mildly relativistic XRBs we can expect that lateral spreading only matters at all when $R>R_{\rm RS}$; we now look to verify this expectation with hydrodynamic simulations. 

\begin{figure}
\centering
\includegraphics[width=\columnwidth]{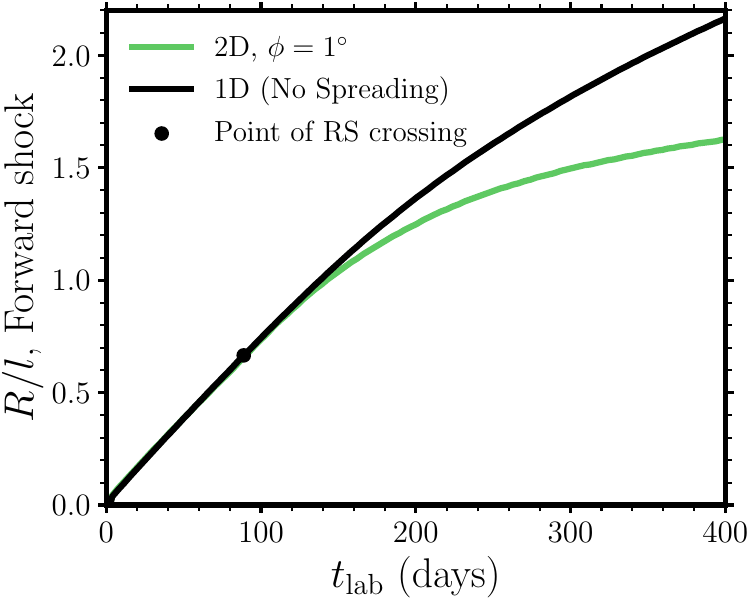}
\caption{The impact of lateral spreading on the forward shock radius, as inferred from a relativistic hydrodynamic simulation. Lateral spreading only occurs after RS crossing. The black line shows the forward shock radius from the 1D (i.e. truly spherical) hydrodynamic simulation described, while the green line shows the forward shock radius from a 2D hydrodynamic simulation of a conical outflow with $\phi = 1^\circ$. Both curves are shown as a function of laboratory frame time, $t_{\rm lab}$, equivalent to our observer frame time at $\theta_i=90^\circ$. Lateral spreading at late times causes more rapid deceleration and leads to a shorter propagation distance. 
}
\label{fig:lateral2}
\end{figure}

\subsection{Lateral spreading with relativistic hydrodynamics}
To estimate the approximate impact of spreading in the XRB case, one could introduce a GRB-like prescription into our swept-up mass parametrisation but only allow it to modify the kinematics when the RS has crossed the shell. In particular, \cite{granot2012} give a simple form of lateral spreading that could be implemented. Instead, we use 2D hydrodynamic simulations in an axisymmetric spherical polar ($r, \varphi$) geometry. 2D relativistic hydrodynamic simulations have been extensively applied to the evolution of highly relativistic blast waves in the GRB-like $\Gamma_0 \gtrsim 10$ regime, where they can be used to, for example, study the dynamics and/or predict afterglow light curves and spectra \citep{zhang2009,vaneerten2010,vaneerten2012,vaneerten2013,duffell2015,granot2018,xie2019,ayache2022,Govreen-Segal2024}. The details of our numerical calculations, which use \textsc{Pluto}, are again described in Appendix~\ref{app:hydro}. The parameters and set up are identical to the 1D calculation used in section~\ref{sec:reverse_shock_time}, except we now introduce an additional parameter in 2D: the half-opening angle $\phi$. 

To investigate spreading in XRBs we conduct 2D simulations with varying values of $\phi$ and identical $\Gamma_0$ and $E_{k,{\rm iso}}$ (see Appendix~\ref{app:hydro} for further details). We first verified that the 1D case produces identical results to 2D when $\phi = 90^\circ$ (corresponding to spherical symmetry). Then, for each simulation, we calculate the position of the forward shock over time, and compare the 2D calculations to the 1D case. The results are shown in Fig.~\ref{fig:lateral2}, which shows $R$ as a function of $t$ from each simulation. First, we do indeed confirm our assertion that the shell does not spread laterally during the RS crossing phase. However, we also find that after RS crossing, the lateral spreading can have a significant impact on the propagation of the shell, causing it decelerate more quickly. The detailed propagation will likely depend on input parameters, and merits further investigation with dedicated hydrodynamics simulations; nevertheless, the basic result -- that lateral spreading only impacts propagation at $R>R_{\rm RS}$, but can cause the shell to propagate less far -- broadly agrees with expectations from GRB theory and general physical reasoning.  

If lateral spreading can cause faster deceleration, it follows that this could impact energy estimates of XRB ejecta, reasoning along similar lines to the reverse shock argument in section~\ref{sec:reverse_shock_time}. Shells that spread laterally will require correspondingly higher energies to produce a given deceleration curve, compared to those that don't. Again, this reasoning suggest that the currently inferred effective energies may be somewhat underestimated, showing how it is important to account for the detailed hydrodynamics of the blast wave. Finally we note two interesting points. One is that XRBs actually do have some constraints on spreading, since their opening angles (or more accurately, the opening angles associated with the observed radio emission) can be estimated from spatially resolved observations, assuming that this emitting region tracks the true dynamics of the ejecta fairly well. The second point is that constraints on lateral spreading -- either from modelling of the displacement over time, or from the radio images directly -- should provide useful (model-dependent) information about initial Lorentz factors: ejecta with modest lateral spreading effects are more likely to have relatively low $\Gamma_0$. 

\section{Discussion}
\label{sec:discuss}
\subsection{Slowing distances and ambient densities}
\label{sec:discuss_slowing}
An approximate Sedov-Taylor-von Neumann distance often quoted for GRBs is $l\sim 10^{18}~{\rm cm}$ \citep[e.g.][]{sari_piran_1995,piran1999}, which is comparable with the {\em observed} deprojected slowing distances for XRBs \citep[e.g.][]{Carotenuto_2024} despite perhaps seven orders of magnitude difference in energetics. However, the estimated value of $l$ (or equivalently, the effective energy $\bar{E}_0$) depends significantly on the ambient density of the medium. 

Direct measurements of GRB jet deceleration or stopping distances are technically challenging, and therefore rare. In order to measure the distance travelled by a jet both VLBI observations (to measure the projected source size) and multi-wavelength modelling of the afterglow (to measure the opening angle) are needed. As a result, there are only about three events where we can say anything with relative confidence about the distance travelled by, and therefore stopping distance of, the jet. These events are GRBs 030329, 130427a and 221009A \citep{2012ApJ...759....4M, 2016MNRAS.462.1111D,2024A&A...690A..74G,rhodes2024}, whose proximity and intrinsic luminosity allowed them to be studied out to much later times than the average event. For both 030329a and 221009A, the combination of VLBI and multi-frequency afterglow observations sets the minimum distance travelled by the jet to be about $4$\,pc \citep{2012ApJ...759....4M,rhodes2024}. We note that at the time of writing, GRB 221009A is still being monitored and so the minimum distance will continue to grow. In the case of GRB 130427a, the minimum distance travelled was measured to be at least 50\,pc \citep{2016MNRAS.462.1111D}. It is thus clear that GRB blast waves can travel significantly further than the canonical $l\sim 10^{18}~{\rm cm}$ figure. 

The circumburst (ambient) densities of GRB environments are inferred from afterglow models. The density profile can be extracted from afterglow light curves and its normalisation is derived from spectral modelling. Is it expected that long GRBs, those produced by collapsing massive stars, have a circumburst density profile following $\rho \propto r^{-2}$ as a result of the progenitor star's stellar wind. Conversely, short GRBs from binary neutron star mergers, are expected to lie in a homogeneous environment $\rho \propto r^{0}$. However, unlike theory, studies of long GRB afterglows find a broad range of (model-dependent) density profiles \citep[e.g.][who found $\rho \propto r^{-2.2}$ and $\propto r^{0}$, respectively]{2019MNRAS.486.2721B,2020MNRAS.496.3326R}. Furthermore, afterglow modelling attempts also find a large range of number density normalisations for both long and short GRBs. \citet{2015ApJ...815..102F} presented a comprehensive set of short GRB observations and found that their ambient number density normalisations ranged between $10^{-5}~{\rm cm}^{-3}$ and $1~{\rm cm}^{-3}$. \citet{aksulu_exploring_2022} performed a similar analysis for long GRBs and found on the whole the number densities were higher but had just as large a range between $10^{-2.5}~{\rm cm}^{-3}$ and $10^{2.5}~{\rm cm}^{-3}$.

It has been suggested for a number of years that XRBs lie in under-dense environments relative to the canonical ISM density of $n \sim 1~{\rm cm}^{-3}$ \citep{heinz2002,wang_external_2003}. This finding appears to be confirmed from kinematic modelling of more recent ThunderKAT sources, with \cite{Carotenuto_2024} finding high effective energies and favouring low ambient medium densities as a way to avoid restrictively high jet powers (under the assumption that the jet launching timescale is roughly the radio flare timescale). This conclusion is broadly supported by other authors. \citesavard\ show that simulations of XRB jet ejecta also require a low density ISM and that the jet ejection itself can create a low-density cavity. Thus, unlike the GRB case, there appears to be an emerging consensus that XRB jet environments are systematically underdense, with $ n \lesssim 10^{-3}~{\rm cm}^{-3}$.

With these observational constraints in mind we can then revisit the apparent coincidence in stopping distances or Sedov-Taylor-von Neumann lengths. 
In some cases, it is relatively likely that GRB blast waves travel a similar distance to XRB discrete jet ejecta before slowing to non-relativistic velocities. For this to be true, the larger energy (factor of $\sim 10^7$) needs to be compensated by a corresponding factor in $\phi^2 \rho_0$, which is feasible if the GRB lives in a dense environment and/or its outflow has a large opening angle (there are indeed GRBs with inferred opening angles significantly larger than a few degrees; \citealt{Golstein2016}). However, in general, GRBs likely have a large range of slowing distances as expected from their range of inferred $E_{k,{\rm iso}}$ values and circumburst densities, and the $l\sim 10^{18}~{\rm cm}$ figure is far from universal.

\begin{figure}
\centering
\includegraphics[width=\linewidth]{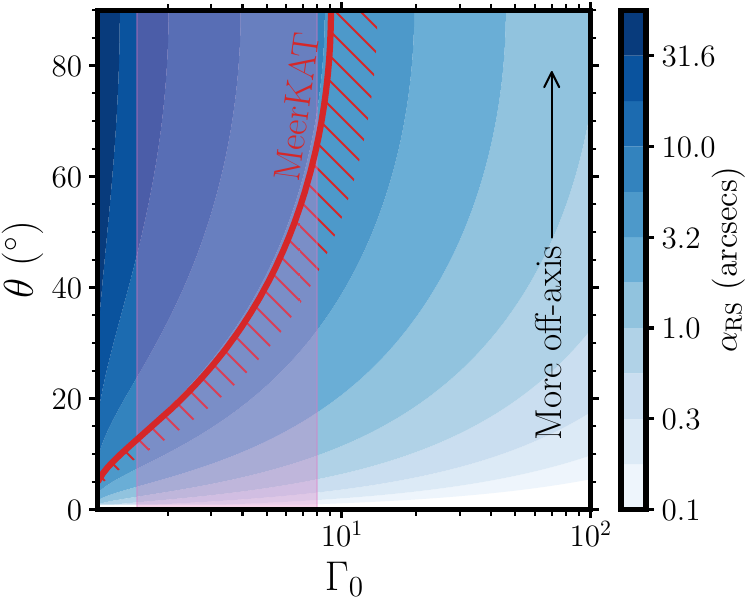}
\caption{Contours of the angular separation at the time of reverse shock crossing, $\alpha_{\rm RS}$, as a function of $\Gamma_0$ and $\theta$ for a source with our representative XRB parameters at a distance of $D=3$~kpc. The angular separation is calculated directly from equation~\ref{eq:alpha_RS} and the red line marks a MeerKAT nominal angular resolution of $5.4^{\prime\prime}$, with the hashes showing the side of the line where $\alpha_{\rm RS} < 5.4^{\prime\prime}$. The pink band shows the approximate initial Lorentz factor range for XRBs.
}
\label{fig:alpha_RS}
\end{figure}

\subsection{XRBs as shock and particle acceleration laboratories}
\label{sec:discuss_shocks}
If indeed $\Gamma_0 \sim{\rm a~few}$, the property that $l / R_{\rm RS} \sim {\rm a~few}$ and $t_{\rm RS}\sim$ tens of days makes XRBs {\em unique shock laboratories}, for a number of reasons. The most obvious one is that their reverse shocks persist for a long time, compared to on-axis GRBs (Fig.~\ref{fig:times}). This makes it much more practical to obtain multi-wavelength coverage, particularly in the radio, where it might take a few days or more to get on source. This advantage is made more pronounced by the fact that the receding ejecta can also be detected in many XRBs, and the associated observed timescales are even longer for the receding case (see section~\ref{sec:receding}).
In addition, the reverse shock persists through a phase when the ejection is a long way from the nucleus. To illustrate this latter point, we can define the angular separation at RS crossing, which at $z=0$ is given by 
\begin{equation}
\alpha_{RS} = D^{-1} R_{\rm RS} \sin\theta = D^{-1} \sin \theta \left (\frac{3 E_0}{\Omega \rho_0 \Gamma_0 (\Gamma_0 - 1)} \right)^{1/3} \, .
\label{eq:alpha_RS} 
\end{equation}
We plot $\alpha_{RS}$ as a filled contour as a function of $\Gamma_0$ and $\theta$ in Fig.~\ref{fig:alpha_RS}. The figure shows that off-axis, moderately relativistic Galactic sources can have their separations from the core at $t_{\rm RS}$ resolved with present day radio telescopes without the need for VLBI. 

There are also intrinsic properties of the shocks in XRBs that make them interesting testbeds; to see this, we must first review the theory of particle acceleration at (trans-)relativistic shocks. Arguably the two most important parameters governing particle acceleration at shocks are the Lorentz factor, $\Gamma_s$, and the magnetisation. 
Both of these parameters affect the efficiency of particle acceleration (the fraction of the total energy given to nonthermal particles), the maximum individual energies those particles reach, and the shape or slope of the particle spectrum \cite[see][for reviews]{kirk1999,bykov2012,marcowith_microphysics_2016,matthews_particle_2020}. Efficient particle acceleration occurs when particles cross the shock many times. At low magnetisations, relativistic shocks are thought to be efficient particle accelerators able to channel a significant fraction of the shock power into nonthermal particles \citep{spitkovsky_particle_2008,sironi_particle_2009,sironi_particle_2011}. At high magnetisation, the efficiency of particle acceleration depends significantly on the magnetic field orientation and is often inefficient \citep[e.g.][]{begelman1990,gallant1992,sironi_particle_2009,sironi_particle_2011}, with the particles unable to `outrun' the shock before being swept downstream. Irrespective of the magnetisation, relativistic shocks are quasi-perpendicular and have steep particle spectra compared to non-relativistic shocks \citep{kirk_particle_1998,achterberg_particle_2001}; both these effects conspire to prevent the driving of strong Larmor-scale turbulence at the highest energies and limit the maximum energy significantly \citep{lemoine2010,sironi_maximum_2013,reville_maximum_2014,bell_cosmic_2019}. However, much of the physics remains an area of theoretical debate, and so having good astrophysical laboratories to test these theories is essential. Furthermore, in the most general sense, we should expect all the parameters that describe the distribution of accelerated particles to depend on the shock Lorentz factor. 

As long as the ejecta shell is denser than the surroundings, the Lorentz factor of the reverse shock is lower than that of the forward shock. Depending on the initial Lorentz factor, and which shock dominates the particle acceleration, the shock will transition between different regimes of Lorentz factor/shock velocity, and in turn sample different regimes of particle acceleration physics. This is the case even if the shock in question is never ultra-relativistic. For example, \cite{bell2011} show that even at modest velocities of $10,000~{\rm km~s}^{-1}$, there are already higher order anisotropies in the particle distribution function that cause the diffusion approximation to break down, with a knock-on impact on particle spectral indices. In addition, \cite{bell_cosmic_2019} show that spectral steepening due to energy exchange between MHD turbulence and the cosmic rays can occur at non-relativistic, shock velocities. Thus regardless of the precise shock velocity, the ability to track a particle accelerator spatially and spectrally -- in a time-resolved manner as the shock transitions between different velocity regimes -- is likely to be a powerful probe of particle acceleration physics. In addition, future combined kinematic-radiative modelling of XRB large-scale jet ejecta \citep{cooper_reverse} can provide constraints on the same microphysical parameters ($\epsilon_e$ and $\epsilon_b$) use in GRB model fitting, allowing tests of particle acceleration theory in a new regime. 

The importance of XRB jets/ejecta as high-energy particle accelerators is emphasized by recent very high energy (VHE) gamma-ray results. The High-Altitude Water Cherenkov observatory (HAWC) has reported extended VHE gamma-ray emission from both SS 433 \citep{Abeysekara2018} and V4641 Sgr \citep{alfaro2024}. In addition, the Large High Altitude Air Shower Observatory (LHAASO) detected five XRBs in $>100$~TeV gamma-rays \citep{lhaaso2024}. Intriguingly, the LHAASO detections included VHE emission aligned with the receding jet of MAXI J1820$+$070, a source now famous for its long-lived superluminal jet ejections \citep{bright_extremely_2020}. As discussed also by \citesavard, these results, particularly the MAXI J1820$+$070 detection, provide the tantalising possibility that XRB large-scale jets are accelerating particles to PeV energies, with the forward and reverse shocks both being candidate acceleration sites. We will explore this possibility further, including the potential cosmic ray and neutrino contributions, in a forthcoming paper (Bacon et al., in prep.). 

\subsection{Constraining the initial Lorentz factor}
\label{sec:lorentz_discuss}

The ratio of the Sedov-Taylor-von Neumann length to the reverse shock crossing time depends only on $\Gamma_0$ (Equation~\ref{eq:lorentz_ratio}). This fact has been used, when estimates of the total burst energy, circumburst density and reverse shock crossing time are available, to infer initial Lorentz factors in GRBs \citep[e.g.][]{laskar2019}. The same principle has also been discussed by \cite{Generozov2017} in the context of TDEs. Using this $R_{\rm RS}/l$ ratio to constraint $\Gamma_0$ in XRBs should also be possible, where there are two big advantages. First, it is significantly easier to estimate $l$, because the trajectory is resolved. Second, as we have argued in the previous subsection, the reverse shock can persist for tens of days. However, although other authors have argued for reverse shock emission in XRB jet ejections \citep[][\citealtsavard]{wang_external_2003}, there is no definitive detection of reverse shock emission (nor for that matter, forward shock emission) -- that is to say, the site of {\sl in situ} particle acceleration is not yet well constrained. 

If a reverse shock component can be clearly identified, there is the inviting possibility that its crossing time can be used to measure the Lorentz factor in XRBs. Furthermore, rather than taking a simple ratio, more progress can be made with a full Bayesian fitting procedure that incorporates both angular separation and light curve data for the ejecta and fits the kinematics and radiation simultaneously. We have taken the first steps towards this in a companion paper \citep{cooper_reverse}. 

\section{Summary and Conclusions}
\label{sec:conclusions}

We have revisited blast wave modelling in GRBs and XRBs and examined the differences and similarities between these different classes of source. We started by providing the observational context and described the various selection effects and biases at work when placing XRBs and GRBs in regimes of energy and Lorentz factor. We then reviewed blast wave models in both the ultra-relativistic, on-axis regime and the more general transrelativistic, off-axis regime, before deriving various critical radii and times during the blast wave evolution. Our main findings are as follows: 

\begin{itemize}
    \item In section~\ref{sec:blast-wave-models}, we demonstrate and emphasize that XRBs are the off-axis, moderately relativistic cousins of GRBs (e.g. Fig.~\ref{fig:dR_dt}); although the Blandford-Mckee model and the $R \propto \Gamma^2 t$ behaviour are specific to GRBs, the more general equations for $dR/dt$ and energy conservation can be applied in both regimes, as originally suggested by \cite{Huang_1999}.
    \item We show that, in moderately relativistic XRB jet ejecta, the RS crossing phase is long-lived and lasts for tens of days in the observer frame (Fig.~\ref{fig:times}) as also found in hydrodynamic simulations \citepsavard. The ratio of the blast wave radius at the point of RS crossing to the Sedov-Taylor-von Neumann length is significantly larger ($l/R_{\rm RS}\sim {\rm a~few}$) in moderately relativistic blast waves compared to ultra-relativistic ones (Fig.~\ref{fig:critical_radii}). 
    \item We describe the overall evolution of blast waves in the ultra-relativistic on-axis (GRB) and moderately relativistic off-axis (XRB) cases (Fig.~\ref{fig:schematic} and Fig.~\ref{fig:schematic_log}). The GRB case is well described in the literature, and is characterised by a short RS crossing phase, a relatively long Blandford-Mckee phase and a slow transition to a Newtonian regime. By contrast, the XRB blast wave never enters a true Blandford-Mckee phase and has a long RS crossing -- after the RS crosses it instead enters a pseudo-Blandford-Mckee phase before transitioning to a Sedov-Taylor-von Neumann solution. 
    \item We examine the effect of geometry on XRB ejecta or blob propagation. We discuss the relationship between relativistic `cloud-crushing' and the quasi-spherical case, which both produce similar results (section~\ref{sec:discuss_cloud}). 
    \item We also explore the impact of lateral spreading on the blast wave, in which $\phi$ increases over time (section~\ref{sec:discuss_lateral}). We argue that the real lateral spreading must depend on the hydrodynamics of the system and applying a GRB recipe during the RS phase is unrealistic -- we therefore argue for a more modest impact of lateral spreading as inferred from our 2D RHD simulations. Nevertheless, any lateral spreading will act to increase energy estimates from kinematic modelling. 
    \item We use our relativistic hydrodynamic simulation to show that if the emission from XRB ejecta comes from particles accelerated at the reverse shock then the propagation distance can be significantly shorter compared to a forward shock scenario (Fig.~\ref{fig:shock_radii}). This effect could lead to an underestimate of the effective energy of the blast wave if a forward shock model is then applied. 
    \item We argue that discrete jet ejecta from XRBs are excellent laboratories for studying reverse shocks and particle acceleration in moderately relativistic shocks (section~\ref{sec:discuss_shocks}). This is because their reverse shock is long-lived (Fig.~\ref{fig:times})  and crosses the ejecta at a point when the separation can be easily resolved with present day radio telescopes (Fig.~\ref{fig:alpha_RS}).
    \item If RS signatures can be robustly identified in XRBs, then the ratio $l/R_{\rm RS}$ can in principle be used to constrain the initial Lorentz factor, $\Gamma_0$, as has been suggested for GRBs. Moving forward, a combined radiative-kinematic modelling framework \citep{cooper_reverse} should allow parameters of the system (such as $\Gamma_0$) to be constrained more rigorously and with fewer degeneracies. 
\end{itemize}
Overall, our work emphasizes that XRB and GRB blast waves and/or jets are complementary laboratories. Each can inform the other, and each probes different regimes in terms of Lorentz factor, energetics, viewing angles and microphysics. XRBs have one big advantage in that their trajectories can be tracked much more easily without the need for VLBI. This property can be exploited so as to learn more about the physics of particle acceleration, shocks and ISM interaction in relativistic plasmas, making an even firmer case for continued targeted, high-resolution radio and X-ray monitoring of large-scale jet ejecta from X-ray binaries.  In turn, improved constraints on energetics and Lorentz factors of the ejecta in both XRBs and GRBs can help us understand the central accreting compact object that ultimately powers these extreme transient phenomena. 

\section*{Data Availability}
The bulk of the data created for this article and code for creating most of the figures can be found in a public github repository at \url{https://github.com/jhmatthews/blastwave}, with an associated Zenodo record and digital object identifier (\href{https://doi.org/10.5281/zenodo.15011341}{10.5281/zenodo.15011341}). The Lorentz factor data for X-ray binaries shown in Fig.~\ref{fig:lorentz} will be published by Fender \& Motta (submitted). Any other data will be made available upon reasonable request. 

\section*{Acknowledgements}
We thank the anonymous referee for a helpful and constructive report. We would like to thank Gavin Lamb, Tony Bell, and the ThunderKAT and XKAT collaborations for many helpful discussions. JHM and EE acknowledge funding from a Royal Society University Research Fellowship (URF\textbackslash R1\textbackslash221062). AJC acknowledges support from the Oxford Hintze Centre for Astrophysical Surveys which is funded through generous support from the Hintze Family Charitable Foundation. LR acknowledges support from the Canada Excellence Research Chair in Transient Astrophysics (CERC-2022-00009). KS acknowledges support from the Clarendon Scholarship Program at the University of Oxford and the Lester B. Pearson Studentship at St John's College, Oxford. RF gratefully acknowledges support from UKRI, The Hintze family charitable foundation, and the ERC. SEM acknowledges support from the INAF Fundamental Research Grant (2022) EJECTA. We gratefully acknowledge the use of the following software packages: astropy \citep{astropy-collaboration13,astropy-collaboration18}, matplotlib \citep{matplotlib}, PLUTO \citep{mignone_pluto_2007}.

\input{main.bbl}

\appendix 

\section{Lorentz factor estimates for GRBs}
\label{sec:grb_table}

In Table~\ref{grb_table1} (\ref{grb_table2}) we give the initial Lorentz factor estimates (lower-limits) for the GRBs used in Fig.~\ref{fig:lorentz}.

\begin{table}
    \centering
    \begin{tabular}{ccc}
        \hline
        GRB ID & $\Gamma_0$ estimate & Reference \\
        \hline
        080210a & 38.3 &  \cite{Xi2017} \\
        080810a & 43.7 &  \cite{Xi2017} \\
        080810a & 80.7 &  \cite{Xi2017} \\
        080928a & 29.2 &  \cite{Xi2017} \\
        080928a & 24.3 &  \cite{Xi2017} \\
        081008a & 55.8 &  \cite{Xi2017} \\
        100906a & 49.4 &  \cite{Xi2017} \\
        110801a & 29.2 &  \cite{Xi2017} \\
        110801a & 62.7 &  \cite{Xi2017} \\
        12102a  & 17.1 &  \cite{Xi2017} \\
        121024a & 23.0 &  \cite{Xi2017} \\
        121211a & 34.2 &  \cite{Xi2017} \\
        130427b & 33.4 &  \cite{Xi2017} \\
        130606a & 37.7 &  \cite{Xi2017} \\
        130606a & 19.6 &  \cite{Xi2017} \\
        130606a & 44.9 &  \cite{Xi2017} \\
        130606a & 106.6 &  \cite{Xi2017} \\
        130606a & 39.6 &  \cite{Xi2017} \\
        140512a & 28.6 &  \cite{Xi2017} \\
        140515a & 87.2 &  \cite{Xi2017} \\
        160509a & 330 &  \cite{laskar2016} \\
        130427a & 130 &  \cite{laskar2013} \\
        100219a & 260 &  \cite{Mao2012}\\
        081029a & 500 &  \cite{Holland2012} \\
        081029a & 120 &  \cite{greiner2009} \\
        140629a & 82  &  \cite{hu2019} \\
        181201a & 103 &  \cite{laskar2019} \\
        160625b & 58  &  \cite{Lin2019} \\
        080319b & 520 &  \cite{Fraija2018} \\
        130427a & 550 &  \cite{Fraija2018} \\
        201015a & 204 &  \cite{Ror2022} \\
        201216c & 310 &  \cite{Ror2022}  \\
        110731a & 580 &  \cite{Lu2017} \\
        110731a & 154 &  \cite{Lu2017} \\
        070208a & 115 &  \cite{Liang2010} \\
        080319c & 301 &  \cite{Liang2010} \\
        990123a & 966 &  \cite{Liang2010} \\
        050922c & 401 &  \cite{Liang2010} \\
        060210a & 381 &  \cite{Liang2010} \\
        071010b & 309 &  \cite{Liang2010} \\
        071112c & 244 &  \cite{Liang2010} \\
        \hline
    \end{tabular}
    \caption{Estimates of the initial Lorentz factor in GRBs, with references.}
    \label{grb_table1}
\end{table}

\begin{table}
    \centering
    \begin{tabular}{ccc}
        \hline
        GRB ID & $\Gamma_0$ limit & Reference \\
        \hline
        051111a & 395 &  \cite{Liang2010} \\
        080319b & 486 &  \cite{Liang2010} \\
        060908a & 455 &  \cite{Liang2010} \\
        060912a & 307 &  \cite{Liang2010} \\
        061021a & 363 &  \cite{Liang2010} \\
        071003a & 483 &  \cite{Liang2010} \\
        021211a & 282 &  \cite{Liang2010} \\
        050319a & 337 &  \cite{Liang2010} \\
        050525a & 384 &  \cite{Liang2010} \\
        160625B & 100 &  \cite{alexander2017} \\
        190114c & 140 &  \cite{acciari2019} \\
        191016a & 90  &  \cite{Smith2021} \\
        060124a & 5.3  &  \cite{Yi2015} \\
        060124a & 27.2 &  \cite{Yi2015} \\
        060210a & 45.9 &  \cite{Yi2015} \\
        060210a & 63.3 &  \cite{Yi2015} \\
        060418a & 62.2 &  \cite{Yi2015} \\
        060526a & 12.3 &  \cite{Yi2015} \\
        060526a & 20   &  \cite{Yi2015} \\
        060526a & 17.9 &  \cite{Yi2015} \\
        060526a & 29.1 &  \cite{Yi2015} \\
        060707a & 12.2 &  \cite{Yi2015} \\
        060714a & 68   &  \cite{Yi2015} \\
        060714a & 30.8 &  \cite{Yi2015} \\
        060714a & 44.4 &  \cite{Yi2015} \\
        060714a & 37.3 &  \cite{Yi2015} \\
        060729a & 4.5  &  \cite{Yi2015} \\
        060814a & 28.1 &  \cite{Yi2015} \\
        060906a & 56   &  \cite{Yi2015} \\
        060906a & 80.1 &  \cite{Yi2015} \\
        070306a & 25.3 &  \cite{Yi2015} \\
        070318a & 8.8  &  \cite{Yi2015} \\
        070318a & 9.3  &  \cite{Yi2015} \\
        070721Ba & 56.5 &  \cite{Yi2015} \\
        070721Ba & 62.9 &  \cite{Yi2015} \\
        070721Ba & 74.6 &  \cite{Yi2015} \\
        071031a & 17.3 &  \cite{Yi2015} \\
        071031a & 27.8 &  \cite{Yi2015} \\
        071031a & 29.8 &  \cite{Yi2015} \\
        071031a & 25.8 &  \cite{Yi2015} \\
        050416A & 43.7 &  \cite{Yi2015} \\
        050802a & 10.3 &  \cite{Yi2015} \\
        050814a & 36.6 &  \cite{Yi2015} \\
        050814a & 20.1 &  \cite{Yi2015} \\
        050820a & 12.3 &  \cite{Yi2015} \\
        050904a & 20.8 &  \cite{Yi2015} \\
        050904a & 30.9 &  \cite{Yi2015} \\
        050904a & 24.7 &  \cite{Yi2015} \\
        050904a & 25.4 &  \cite{Yi2015} \\
        050904a & 55.9 &  \cite{Yi2015} \\
        050904a & 34.9 &  \cite{Yi2015} \\
        050904a & 14.3 &  \cite{Yi2015} \\
        051016b & 15.1 &  \cite{Yi2015} \\
        060115a & 38.7 &  \cite{Yi2015} \\
        \hline
    \end{tabular}
    \caption{Estimated lower limits on the initial Lorentz factor in GRBs.}
    \label{grb_table2}
\end{table}

\begin{figure*}
    \centering
    \includegraphics[width=\linewidth]{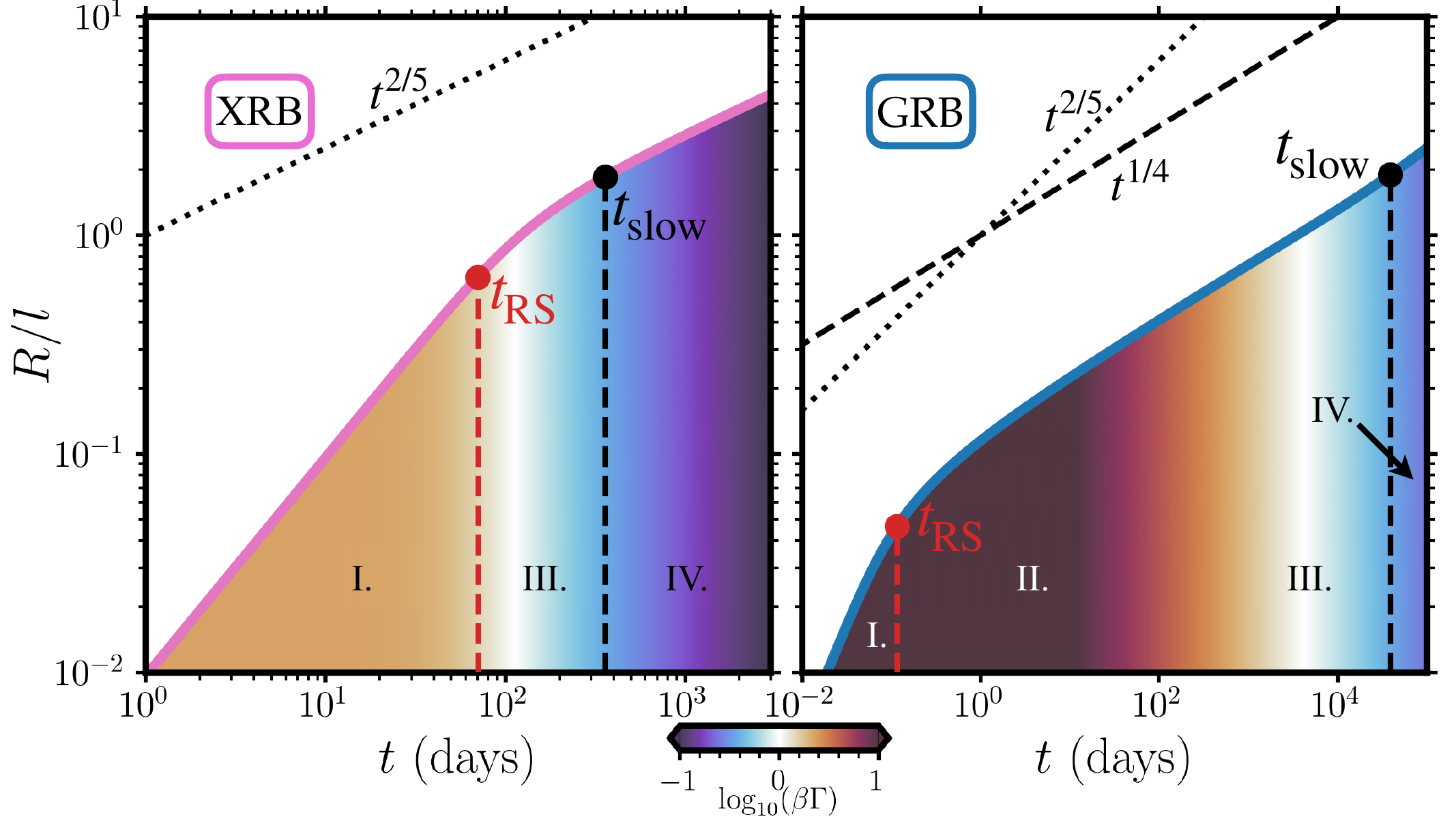}
    \caption{Analogue to Fig.~\ref{fig:schematic}, but with logarithmic axes and expected Sedov-Taylor-von Neumann and \citetalias{blandford_fluid_1976} time evolutions marked with dotted and fashed lines, respectively. The colours and labeled zones I-IV match those denoted in Fig.~\ref{fig:schematic}. 
    }
    \label{fig:schematic_log}
    \end{figure*}

\section{Relativistic Hydrodynamic Simulations}
\label{app:hydro}
We ran 1D and 2D relativistic hydrodynamic (RHD) simulations, using {\sc Pluto} \citep{mignone_pluto_2007}. The aim of the simulations was to test the reverse shock crossing condition, predict reverse shock observational signatures, and examine the impact of lateral spreading and 2D effects. We use {\sc Pluto} to solve the equations of RHD in both 1D spherical and 2D polar axisymmetric ($r,\varphi$) geometries. We initialise the simulations with a shell of lab-frame width $\Delta_0 = 2\times10^{16}~{\rm cm}$, Lorentz factor $\Gamma_0 = 2.5$ and mass $M_0 = 9.74 (\Omega / 4\pi) \times 10^{26}~{\rm g}$ such that the initial rest mass density inside the shell is $\rho_0 = 3 M_0 / (\Omega \Delta_0^3 \Gamma_0) = 9.25\times 10^{26}~{\rm g}$. This results in an isotropic energy equivalent of $E_{k,{\rm iso}} = 1.313 \times 10^{48}~{\rm erg}$, corresponding to an initial kinetic energy of the shell for $\phi = 1^\circ$ of $E_0 = 10^{44}~{\rm erg}$. Both the 1D and 2D setups have resolutions of $10^{15}~{\rm cm}$ and a maximum radius of $1.024\times10^{18}~{\rm cm}$ leading to 1024 radial cells in the 1D case and $1024 \times 512$ ($r,\varphi$) cells in the 2D case. We evolve the simulations using the HLLC Riemann solver \citep{mignone_hllc_2005}, a Courant-Friedrichs-Levy number of 0.4 and linear reconstruction. We adopt the Taub-Mathews \citep{taub_relativistic_1948,mignone_equation_2007} equation of state, the Monotized Central flux limiter and 2nd order Runge-Kutta time stepping. This setup is a standard recommended one for {\sc Pluto} RHD simulations, but we nevertheless experimented with the numerical scheme, trying higher-order methods and various different Riemann solvers and flux limiters. Our results were not sensitive to these choices. We also verified that a 2D simulation with spherical symmetry ($\phi=90^
{\circ}$) produced the same results as the 1D simulation. 

We assigned a tracer fluid ($Q$) to the initial shell material, and evolved this tracer as a passive scalar in the simulation. We then define shell/ejecta material as having $Q>10^{-2}$ and $Q\leq10^{-2}$, respectively. We identified the forward shock by searching inwards for a jump in pressure and use its position to calculate $R(t)$. The reverse shock proxy in Fig.~\ref{fig:shock_radii} is taken to be the maximum distance reached by the ejecta material, i.e. the maximum $r$ at which $Q>10^{-2}$. To find the point of reverse shock crossing, we inspected the 1D spatial profiles of density, pressure, velocity, and tracer, to find the time at which the reverse shock finishes crossing the shell. We found that this does indeed match well with the point at which the swept up mass equals $1/\Gamma_0$ of the initial blob mass, so we record the value of $R$ at which $m_{\rm sw} = M_0 / \Gamma_0$ as found numerically from the simulation results. For this purpose, we calculate the swept up mass by summing over all cells (indexed by $n$) with $r<R$ so that 
\begin{equation}
    m_{\rm sw} = \left( \sum_{n, r_n<R} \rho_n \Gamma_n V_n \right) - M_0 \, ,
\end{equation}
that is the total mass behind the forward shock minus the initial shell mass. 

\section{Additional Schematic Figure}
\label{app:additional}
In Fig.~\ref{fig:schematic_log} we show a modified version of Fig.~\ref{fig:schematic} on logarithmic axes. The shading, timescales, and phases are labelled as in Fig.~\ref{fig:schematic}. The logarithmic axes make it easier to see the early time evolution of the GRB, and also the slopes of the power law evolution of $R(t)$. In particular, in the left-hand XRB panel, we see a transition from an early pseudo-ballistic trajectory to a Sedov-Taylor-von Neumann phase with $R \propto t^{2/5}$ at late times when $R>l$ and $R > R_{\rm slow}$. In the right-hand GRB panel, after an initial ballistic phase we instead recover the Blandford-Mckee solution $R\propto t^{1/4}$ for most of the early time evolution. Again, at late times when $R>l$ and $R > R_{\rm slow}$, the $R \propto t^{2/5}$ Sedov-Taylor-von Neumann evolution is gradually recovered \citep[see][for further discussion of the transition to the non-relativistic regime]{livio2000,zhang2009,peer2012,jetsimpy2024}. Somewhat counter-intuitively, this means the XRB $R(t)$ curve gets shallower at large $t$, whereas the GRB curve steepens, due to relativistic and geometric effects.

\bsp	
\label{lastpage}
\end{document}